\documentclass[useAMS,usenatbib]{mn2e}
\usepackage{graphicx,amssymb, txfonts,pstricks}
\usepackage[dvips]{epsfig}
\usepackage{subfig}
\usepackage{float}

\def\ha{H$\alpha$}
\def\hb{H$\beta$}

\def\oiii{[O\,{\sc iii}]}

\def\sii{[S\,{\sc ii}]}
\def\nii{[N\,{\sc ii}]}

\def\h2{H$_2$}
\def\p1{Paper~I}

\def\kms {$\rm km\,s^{-1}$}

\title[Outflows in the NLR of Bright Seyfert Galaxies]{Outflows in the Narrow Line Region of Bright Seyfert Galaxies - I:  GMOS-IFU Data}

\author[Freitas et al.]
{I. C. Freitas$^{1,2}$\thanks{E-mail:izabelfisica@gmail.com}, R. A. Riffel$^{1}$,
T. Storchi-Bergmann$^{3}$,
M. Elvis$^{4}$,
A. Robinson$^{5}$,
\newauthor
D. M. Crenshaw$^{6}$,
N. M. Nagar$^{7}$,
D. Lena$^{8,9}$,
H. R. Schmitt$^{10}$,
 S. B. Kraemer$^{11}$\\
$^{1}$Universidade Federal de Santa Maria, Centro de Ci\^encias Naturais e Exatas, Departamento de F\'\i sica, 97105-900, Santa Maria, RS, Brazil\\
$^{2}$Universidade Federal de Santa Maria, Col\'egio Polit\'ecnico, 97105-900, Santa Maria, RS, Brazil \\
$^{3}$Universidade Federal do Rio Grande do Sul, Instituto de F\'\i sica, CP 15051, 91501-970, Porto Alegre, RS, Brazil\\
$^{4}$Harvard-Smithsonian Center for Astrophysics, 60 Garden Street, Cambridge, MA 02138, USA\\
$^{5}$School of Physics and Astronomy, Rochester Institute of Technology, 84 Lomb Memorial Drive, Rochester, NY 14623-5603, USA\\
$^{6}$Department of Physics and Astronomy, Georgia State University, Astronomy Offices, 25 Park Place, Suite 605, Atlanta, GA 30303, USA\\
$^{7}$Department of Astronomy, Universidad de Concepci{\'o}n, Casilla 160-C, Concepci{\'o}n, Chile\\
$^{8}$SRON, Netherlands Institute for Space Research, Sorbonnelaan 2, NL-3584 CA Utrecht, the Netherlands\\
$^{9}$Department of Astrophysics/IMAPP, Radboud University, Nijmegen, PO Box 9010, NL-6500 GL Nijmegen, the Netherlands\\
$^{10}$Remote Sensing Division, Naval Research Laboratory, 4555 Overlook Avenue, SW, Washington, DC 20375, USA\\
$^{11}$Institute for Astrophysics and Computational Sciences, Department of Physics, The Catholic University of America, Washington, DC 20064, USA}

\begin{document}


\pagerange{\pageref{firstpage}--\pageref{lastpage}} \pubyear{2011}

\maketitle

\label{firstpage}

\begin{abstract}

We present two-dimensional maps of emission-line fluxes and kinematics, as well as of the stellar kinematics of the central few kpc of five bright nearby Seyfert galaxies -- Mrk\,6, Mrk\,79, Mrk\,348, Mrk\,607 and Mrk\,1058 -- obtained from observations with the Gemini Multi-Object Spectrograph (GMOS) Integral Field Unit (IFU) on the Gemini North Telescope. The data cover the inner 3\farcs5$\times$5\farcs0 -- corresponding to physical scales in the range 0.6$\times$0.9 to  1.5$\times$2.2\,kpc$^2$ -- at a spatial resolution ranging from 110 to 280 pc  with a spectral coverage of 4300 -- 7100\,\AA\  and velocity resolution of $\approx$ 90\,km\,s$^{-1}$. The gas excitation is Seyfert like everywhere but show excitation, but show excitation gradients that are correlated with the gas kinematics, reddening and/or the gas density. 
The gas kinematics show in all cases two components: a rotation one similar to that observed in the stellar velocity field, and an outflow component. In the case of Mrk607, the gas is counter-rotating relative to the stars. Enhanced gas velocity dispersion is observed in association to the outflows according to two patterns:  at the locations of the highest outflow velocities along the ionization axis or perpendicularly to it in a strip centered at the nucleus that we attribute to an equatorial outflow. Bipolar outflows are observed in Mrk\,348 and Mrk\,79, while in Mrk\,1058 only the blueshifted part is clearly observed, while in the cases of Mrk\,6 and Mrk\,607 the geometry of the outflow needs further constraints from modeling to be presented in a forthcoming study, where the mass flow rate and powers will also be obtained.

\end{abstract}


\section{Introduction}

The physical processes that couple the growth of supermassive black holes (SMBH) to their host galaxies - the so-called feeding and feedback processes - occur in the vicinity of the galaxy nucleus (inner $\approx$ 1 kpc) \citep{Hopkins10} when it becomes active due to mass accretion to the SMBH \citep{Ferrarese05,Kormendy13}. The radiation emitted by Active Galactic Nuclei (AGN) works as a flashlight that illuminates and ionizes the gas in the vicinity of the nucleus, forming the Narrow Line Region (NLR). Accretion disk winds \citep{Elvis00,Ciotti10} interact with the gas and produce outflows that are observed in the NLR reaching velocities of hundred of km\,s$^{-1}$ \citep{Das06,Storchi-Bergmann10}. Relativistic jets emanating from the AGN also interact with the gas of the NLR. Both types of outflow produce feedback, which is a necessary ingredient in galaxy evolution models to avoid producing over-massive galaxies \citep{Fabian12}. Inflows have also been observed \citep[e.g.][]{taconi94,Riffel08,Riffel13,ms09,Crenshaw10,allan11,allan14a}. The importance of the NLR stems from the fact that it is spatially resolved
and exhibits strong line emission, allowing the observation of the effects of feeding and feedback occurring in this region.

The first imaging studies of the NLR \citep[e.g.][]{Wilson94} showed fan-shaped regions, supporting the Unified Model \citep[e.g.][]{Antonucci93} and long-slit spectroscopy revealed outflows along the cones \citep{Storchi-Bergmann92,Das06,Crenshaw10}. However, Hubble Space Telescope (HST) [O\,{\sc iii}] images of the NLR of a complete sample based on 60$\mu$m  luminosity revealed that the conical morphology is more the exception than the rule \citep{schmitt03}. A recent long-slit HST Space Telescope Imaging Spectrograph (STIS) study was conducted on a sample of 48 AGN, 35 of which showed extended NLR \citep{Fischer13}. Only 1/3 (12) of that sub-sample have outflowing kinematics, while the remaining are classified as ``ambiguous'' or ``complex''.  

Integral Field Spectroscopy (IFS) using large telescopes is a powerful tool to map the NLR of nearby galaxies \citep[e.g.][]{riffel06,barbosa09,mrk1066-kin,mrk1157,harrison14,davies14,diniz15,lin16,fischer17,ms11,ms17,dominika17,dasilva17,bae17}, as they provide the spatial coverage missed by long-slit spectroscopy at resolutions of a few tens of parsecs. We have recently used the Gemini Multi-Object Spectrograph Integral Field Unit (GMOS-IFU), to study the gas distribution, excitation and kinematics of the inner kiloparsec of a few nearby galaxies, including the Seyfert 2 galaxies NGC\,2110 \citep{allan14b} and NGC\,1386 \citep{Lena14,Lena15} as well as Seyfert 1.8 galaxy NGC\,1365 \citep{Lena16}. In these cases, we found that most of the extended NLR emission has kinematics that can be attributed to gas rotating in the galaxy disk. In addition, outflows are observed within the inner $\approx$ 300 pc for NGC\,2110, being quasi-spherical rather than conical. NGC\,1386 has features that suggest the presence of a bipolar outflow, located within the inner $\approx$ 150 pc. \citet{Lena16} found that there is a fan-shaped outflow in the NGC\,1365, as suggested previously \citep[e.g.][]{edmunds88}.
Similar results are found by studies in the near-infrared,  which suggest that the molecular and ionized gas kinematics of the inner few hundred of parsercs of nearby Seyfert galaxies present a combination of both rotation and in gas outflows \citep[e.g.][]{fischer17,Riffel13}.  
IFS is the best way to constrain the structure and kinematics of the NLR. A complete census of the feeding and feedback processes is still lacking.

In this paper, we use IFS to study the gas distribution and kinematics in the inner kpc of five nearby AGNs: Mrk\,6, Mrk\,79, Mrk\,348, Mrk\,607 and Mrk\,1058. Previous optical IFS is available only for Mrk\,348 of our sample \citep{Stoklasova09}, but with a lower spatial resolution than that presented here.  We present and discuss flux and emission-line ratio maps, as well as maps of the gas velocity fields and velocity dispersion. A detailed analysis of the gas kinematics will be presented in a forthcoming paper. In Section 2 we describe the observations and data reduction procedures. In Section 3 we present the emission-line flux and kinematics maps, that are discussed in Section 4. The conclusions of this work are presented in Section 5.

\section{Observations and Data Reduction}\label{obs}
\begin{table*}
\centering 
\begin{center}
\caption{Observations Log.} 
\vskip 1mm
\begin{tabular}{ l c c c c c c c c c} 
\hline
\\[-0.25cm]
\bf Object &\bf Distance &\bf Nuclear  &\bf Spatial    &\bf Spectral       &\bf Exposure &\bf  log L$_{\rm 2-10\,\rm keV}$ &\bf log L$_{\rm [O\,{\sc III}]}$\rm$^e$ &\bf log L$_{[\rm O\,{\sc III}]}$ &\bf IFU Position\\
\bf 	   &\bf	         &\bf Activity &\bf Resolution &\bf Resolution     &\bf Time	 &\bf                              &\bf                                     &\bf                              &\bf Angle\\
\bf 	   &\bf (Mpc)    &\bf          &\bf (pc)       &\bf (km\,s$^{-1}$) &\bf (s)      &\bf (erg\,s$^{-1})$              &\bf (erg\,s$^{-1})$                     &\bf (erg\,s$^{-1})$              &\bf ($^\circ$)\\
\\[-0.25cm]

\hline
\\[-0.2cm]
Mrk\,6  	& 79.0   & Sy 1.5 & 235	& 90  & 7$\times$810 & 43.0$^a$ & 41.72 & 42.24 & 280	\\ \\[-0.3cm]
Mrk\,79		& 91.6	 & Sy 1   & 280	& 90  & 6$\times$810 & 43.4$^b$ & 41.58 & 41.95 & 73  \\ \\[-0.3cm]
Mrk\,348  	& 63.9   & Sy 2   & 190	& 85  & 6$\times$810 & 42.6$^c$ & 41.26 & 41.91 & 185	\\ \\[-0.3cm]
Mrk\,607	& 36.1   & Sy 2   & 110	& 90  & 7$\times$810 & 40.8$^d$ & 40.44 & 41.02 & 137	\\ \\[-0.3cm]
Mrk\,1058	& 71.8   & Sy 2   & 215 & 85  & 6$\times$810 & -        & 40.49 & 41.23 & 121  \\ \\[-0.3cm]
\\[-0.2cm]	 \hline 
\multicolumn{8}{l}{References: $a$ - \citet{lutz04}; $b$ - \citet{kaspi05}; $c$ - \citet{ueda01}; $d$ - \citet{lamassa11}; $e$ - \citet{schmitt03}.}\\
\end{tabular}
\end{center}
\label{record}
\end{table*}

We have selected our sample from the Seyfert galaxies observed in the HST [O\,{\sc iii}] Snapshot Survey \citep{schmitt03}, that comprises 60/88 of all Seyfert galaxies with z $\leq$ 0.031 from the catalog of warm IRAS sources \citep{deGrijp92}. 
From the 60 galaxies, we selected a sub-sample of 30 according to the following criteria: (1) each galaxy presents extended emission beyond $\approx$ 1\farcs0 from the nucleus; (2) the sub-sample spanned the whole range of AGN luminosities from the sample (39 $\leq$ log (L[O\,{\sc iii}]) $\leq$ 42 erg\,s$^{-1}$). We excluded already well studied objects, such as NGC\,4151 and NGC\,1068, and also targets already observed by members of our groups in previous runs. In this work we present a study using new data for the galaxies Mrk\,6, Mrk\,79, Mrk\,348, Mrk\,607 and Mrk\,1058. 

 We used the GMOS operating in the IFU mode on the Gemini North Telescope. The observations were made from September 30, 2014 to January 12, 2015 under the project GN-2014B-Q-87. We used the IFU on the one-slit mode and two sets of observations were made, one centred at 5700 \,\AA\ and another at 5750 \,\AA, to correct for the effects of the gaps between the GMOS CCDs. The B600 grating was used in combination with the G5307 filter in order to obtain spectra in the range from 4300 \,\AA\ to 7100 \,\AA, that includes the most intense emission-lines from the NLR of AGNs: H$\beta$, [O\,{\sc iii}]\,$\lambda\lambda$4959,5007, [O\,{\sc i}]\,$\lambda$6300, H$\alpha$, [N\,{\sc ii}]\,$\lambda\lambda$6548,83\, and [S\,{\sc ii}]\,$\lambda\lambda$6716,31. The number of individual exposures for each galaxy can be found in Table~1 and the total exposure time ranged from 81 to 95 min.

Data reduction was performed using a series of tasks from the GMOS package developed as part of the GEMINI IRAF package. The reduction process followed the standard procedure of GMOS-IFU data reduction (see Lena 2014)  including bias and sky subtraction, flat-fielding, trimming of the images, wavelength and flux calibration, building of the datacubes, final alignment and combination of individual exposure cubes. The individual datacubes were median combined using the position of the peak of the continuum emission as reference to generate the final datacube for each galaxy.

In Table 1, we present the distance as quoted in NED\footnote{NASA/IPAC  EXTRAGALACTIC  DATABASE available at $http://ned.ipac.caltech.edu$}, nuclear activity class, the spatial and spectral resolutions, the exposure time,  2 -- 10 keV X-ray luminosity, [O\,{\sc iii}] luminosity by \citet{schmitt03}, total [O\,{\sc iii}] luminosity obtained from the our data and IFU position angle (PA). The spectral resolution was estimated by measuring the full width at half-maximum (FWHM) of emission-line profiles of the CuAr calibration lamp, used to wavelength calibrate the data. The angular resolution was estimated as the FWHM of the continuum flux distribution of field stars from the acquisition images.

\section{Measurements}

\subsection{Emission-line profile fitting}\label{fits}

\begin{figure*}
\begin{center}
\includegraphics[scale=0.65]{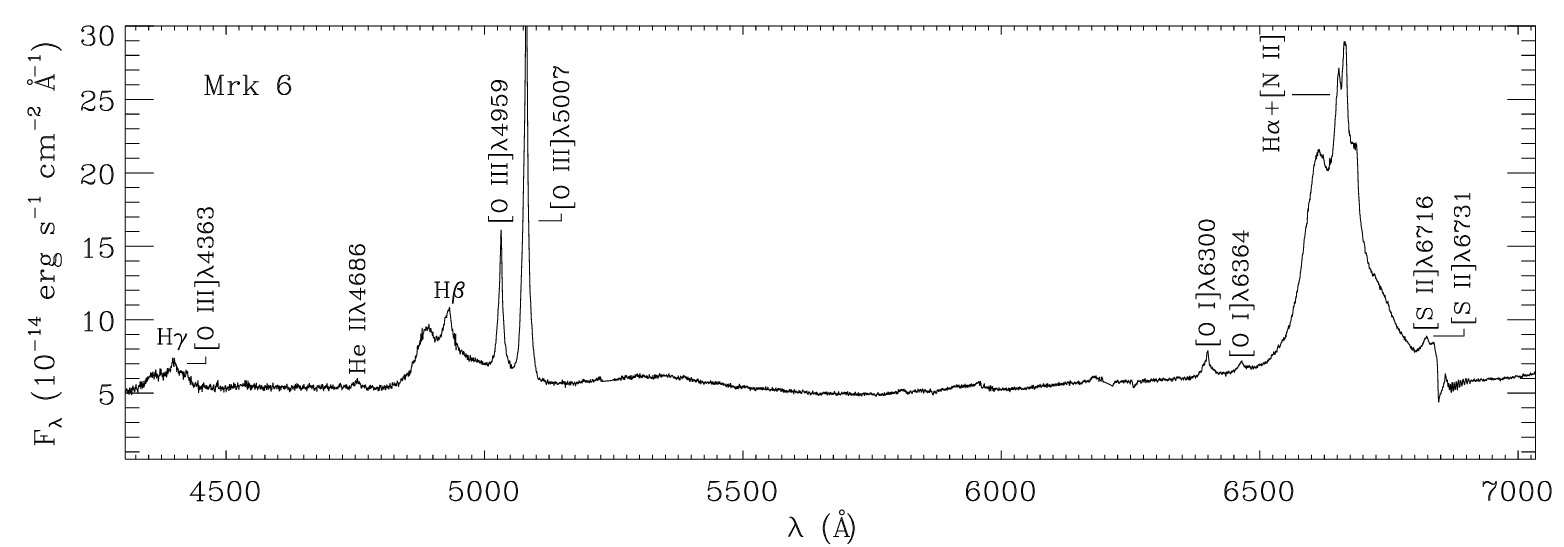}
\includegraphics[scale=0.65]{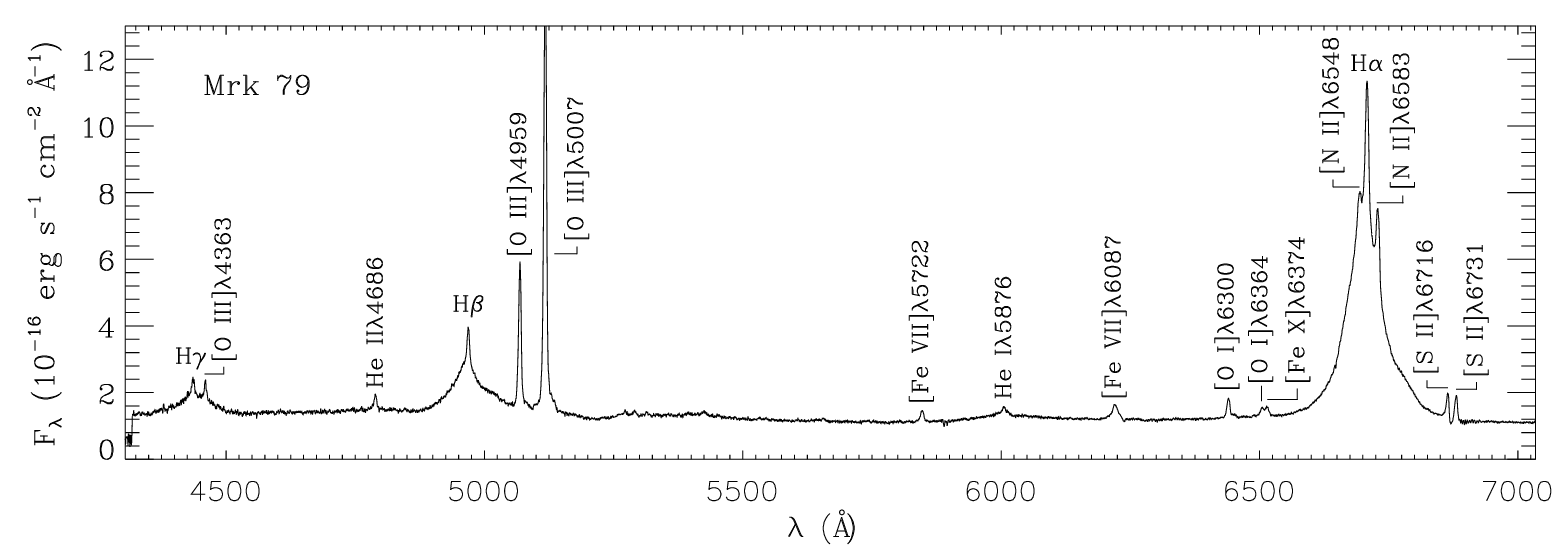}
\includegraphics[scale=0.65]{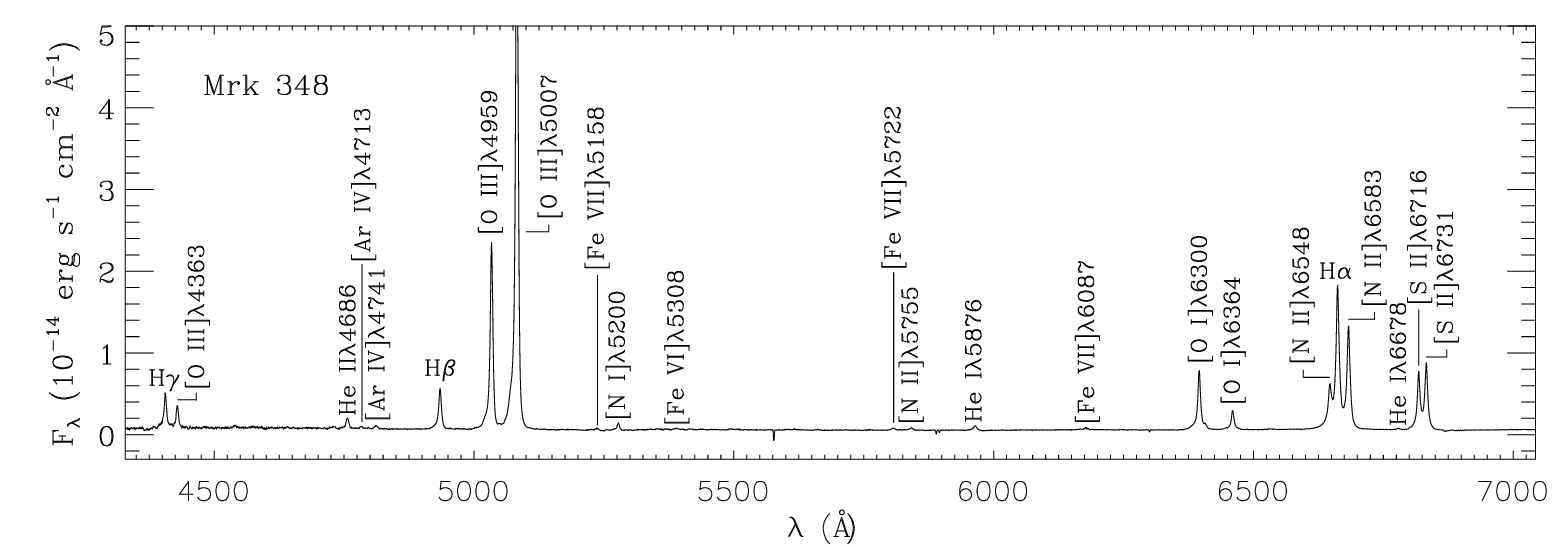}
\includegraphics[scale=0.65]{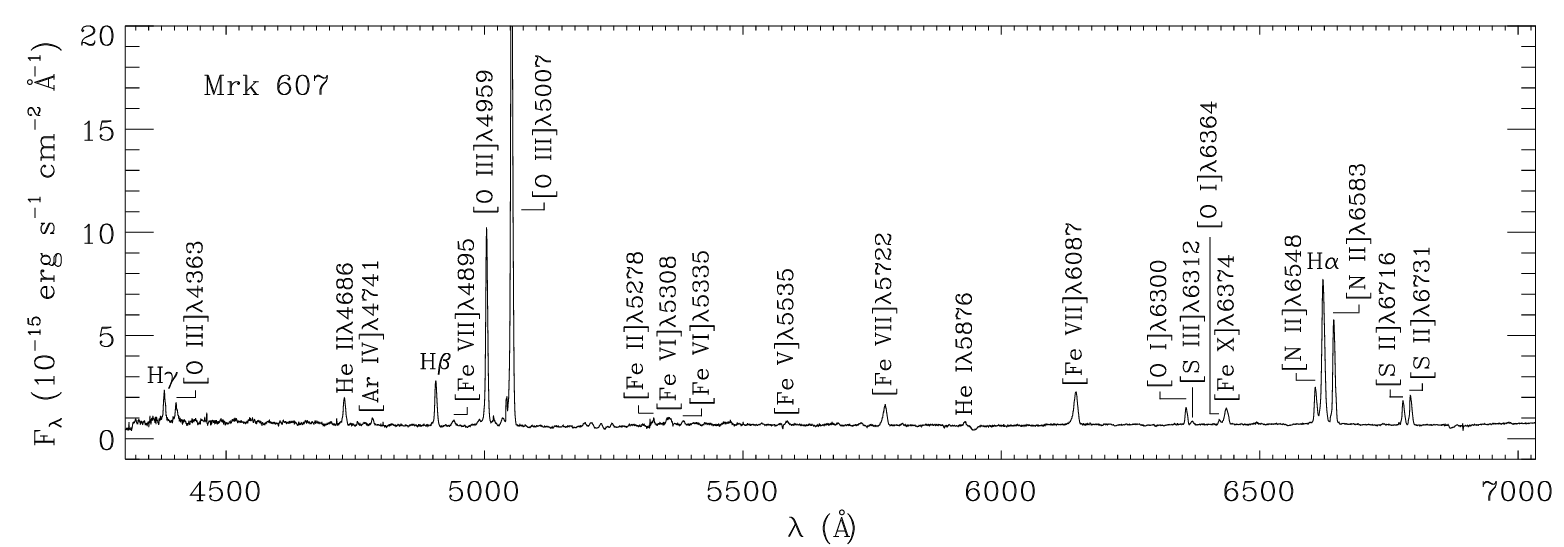}
\includegraphics[scale=0.65]{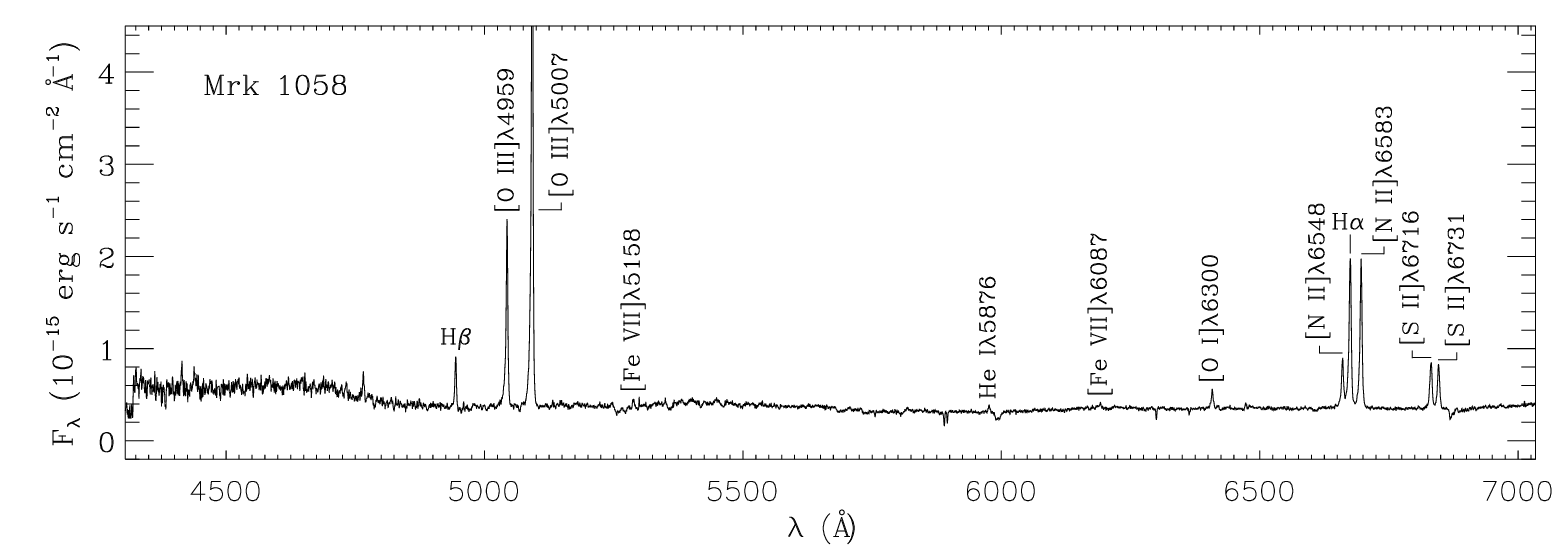}

\caption[spectrum]{  Nuclear spectra of the galaxies of our sample for an aperture of 0\farcs25$\times$0\farcs25. The strongest emission-lines are identified at each panel. From top to bottom: Mrk\,6, Mrk\,79, Mrk\,348, Mrk\,607 and Mrk\,1058.}
\label{spectrum}
\end{center}
\end{figure*}

Figure~\ref{spectrum} presents the nuclear spectra for the five galaxies of our sample, integrated within square apertures of  0\farcs25$\times$0\farcs25, centred at the peak of the continuum emission. The  strongest emission-lines are identified in each spectrum that shows numerous emission-lines (up to about 20).

In order to map the emission-line flux distributions, line-of-sight velocity ($V_{LOS}$) and velocity dispersion ($\sigma$),  we fitted the line profiles at each spaxel by Gaussian curves using  adapted versions of the emission-line PROfile FITting routine  \citep[{profit}][]{Riffel10}, which uses the MPFITFUN routine \citep{mark09}, to perform the non-linear least-squares fit. 

 The following fitting procedure was adopted: 
the [N\,{\sc ii}]$\lambda\lambda6548,83+$H$\alpha$ emission-lines were fitted by keeping the \nii\ flux ratio fixed ([N\,{\sc ii}]$\lambda6583$/[N\,{\sc ii}]$\lambda6548=3$) and tying the central wavelength and width of the [N\,{\sc ii}] lines for all galaxies. For Mrk~6 and Mrk~79, the same procedure was adopted to fit the [O\,{\sc iii}]$\lambda\lambda4959,5007+$H$\beta$ profiles, setting the flux ratio [O\,{\sc iii}]$\lambda5007/$[O\,{\sc iii}]$\lambda4959=3$ and tying kinematics of the [O\,{\sc iii}] lines, while for Mrk\,348 and Mrk\,1058 the H$\beta$ profile was fitted individually and the [O\,{\sc iii}]\,$\lambda$$\lambda$4959,5007\ profiles were fitted simultaneously by using the same constraints above. Finally, for Mrk~607, the [O\,{\sc iii}]  and H$\beta$ lines were fitted individually. These distinct procedures to fit the [O\,{\sc iii}]$\lambda\lambda4959,5007+$H$\beta$ profiles among the galaxies of our sample were adopted in order to have the minimum number of free parameters as possible, resulting in better constrained measurements. The choice of fitting the emission lines individually or simultaneously was done by a visual inspection of the spectra and resulting fits for each galaxy. In all cases, where the emission-line profiles are separated with adjacent continuum regions large enough to constrain its slope, the fit of individual components resulted in better models for the observed profiles.  
The [S\,{\sc ii}] doublet was fitted by keeping the kinematics of the two lines tied, while other emission-lines were fitted individually with all parameters free. 

As the spectral range used to fit each line profile is small, we fitted the underlying continuum by a linear equation.  Although the line profiles at most locations for all galaxies are well reproduced by single gaussian components, double or multiple kinematic components are seen at some locations, in particular for Mrk~348 and Mrk~6. A detailed gas kinematics based on multi-components fits, as well velocity channel maps, will be presented in a forthcoming paper. All maps presented in this paper are based on the fitting of one gaussian for each emission-line profile. In addition, for the Seyfert 1 galaxies Mrk~6 and Mrk~79, we included broad components to fit the H$\alpha$ and H$\beta$ profiles to account for the Broad Line Region (BLR) emission.  Besides the measurement for each parameter, the fitting routine outputs their uncertainties (as estimated by the {\sc mpfitfun} routine) and the resulting $\chi^2$ of the best fit.

\begin{figure*}
\begin{center}
\includegraphics[scale=0.24]{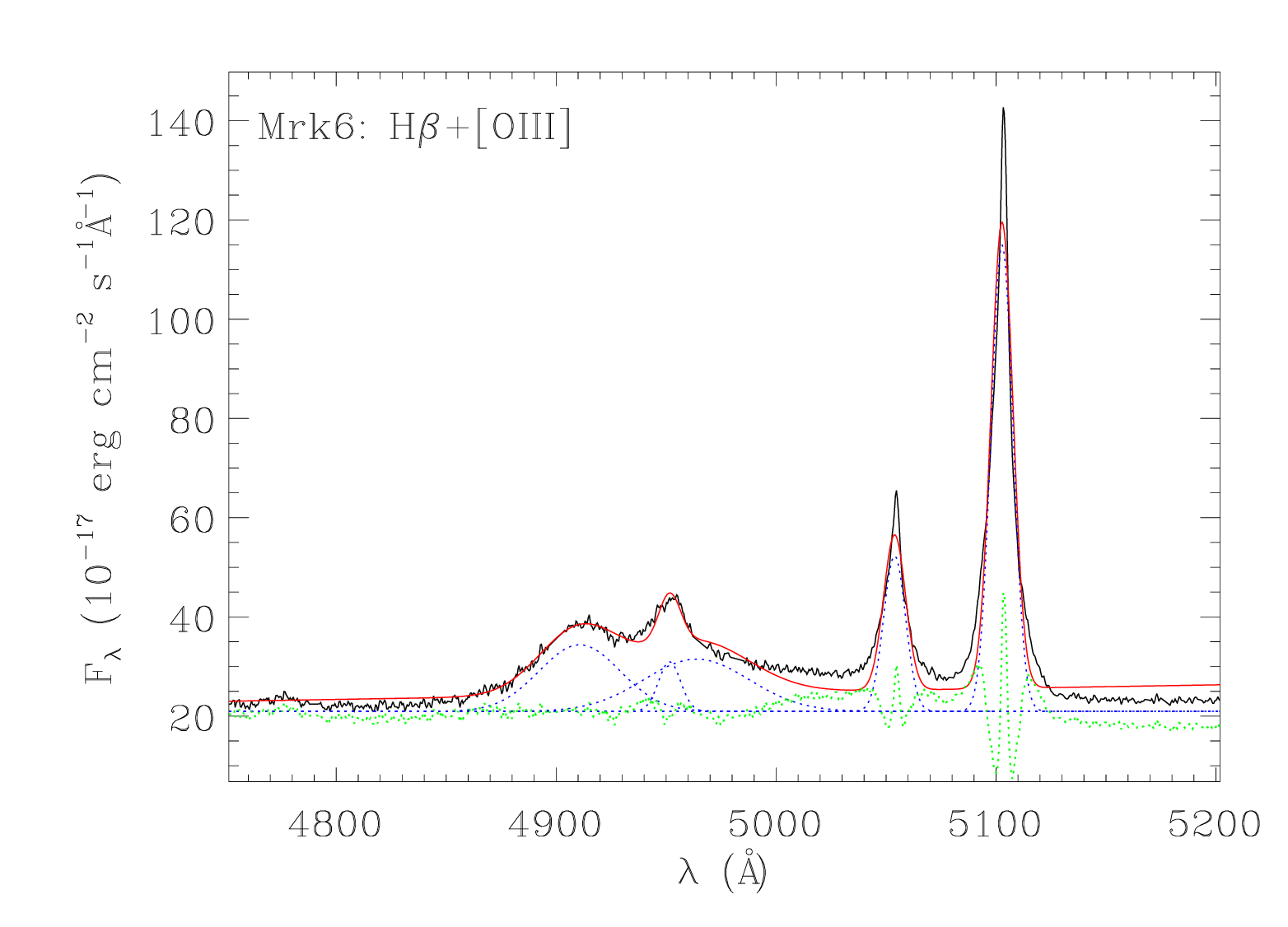}
\includegraphics[scale=0.24]{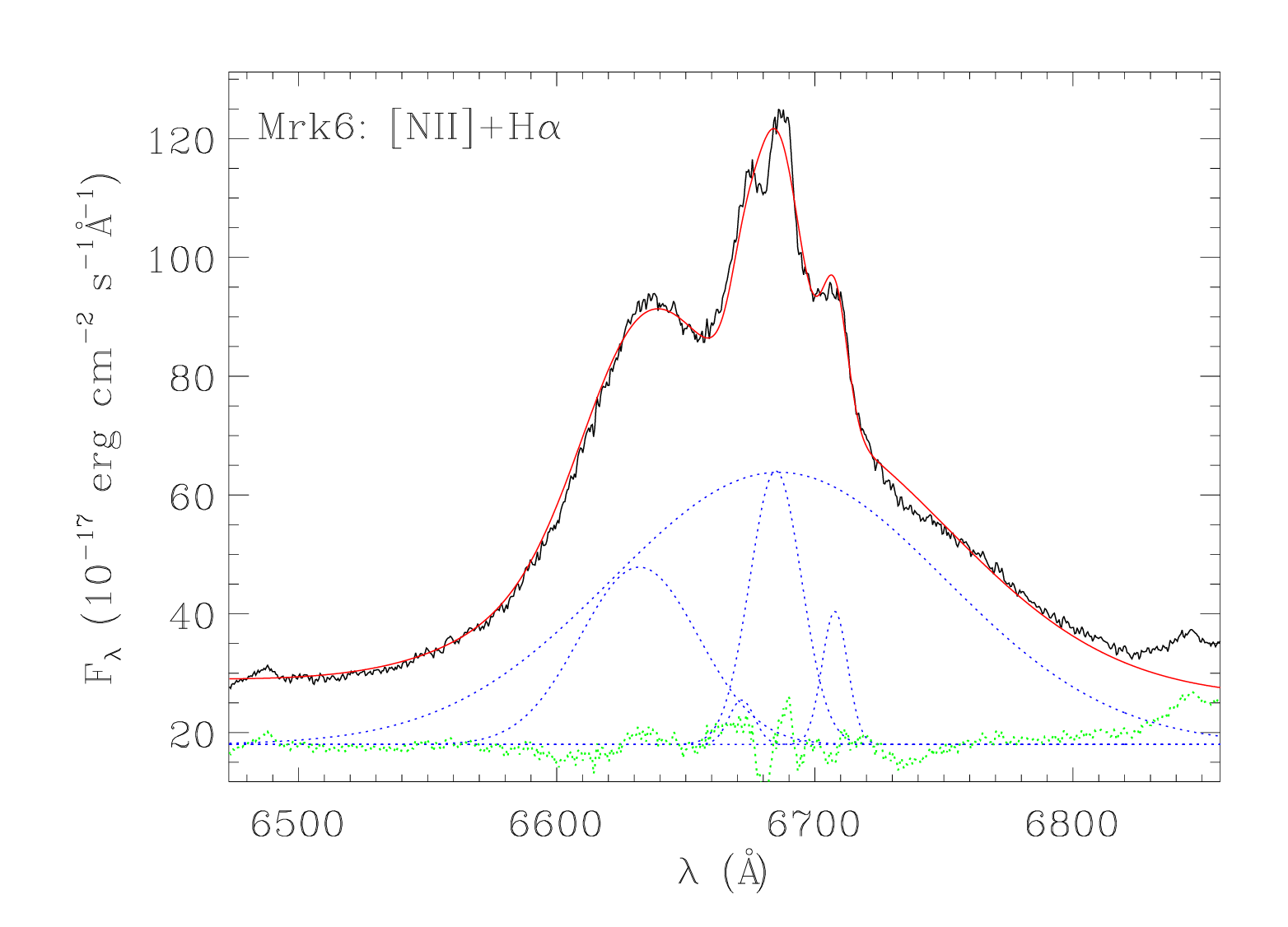}
\includegraphics[scale=0.24]{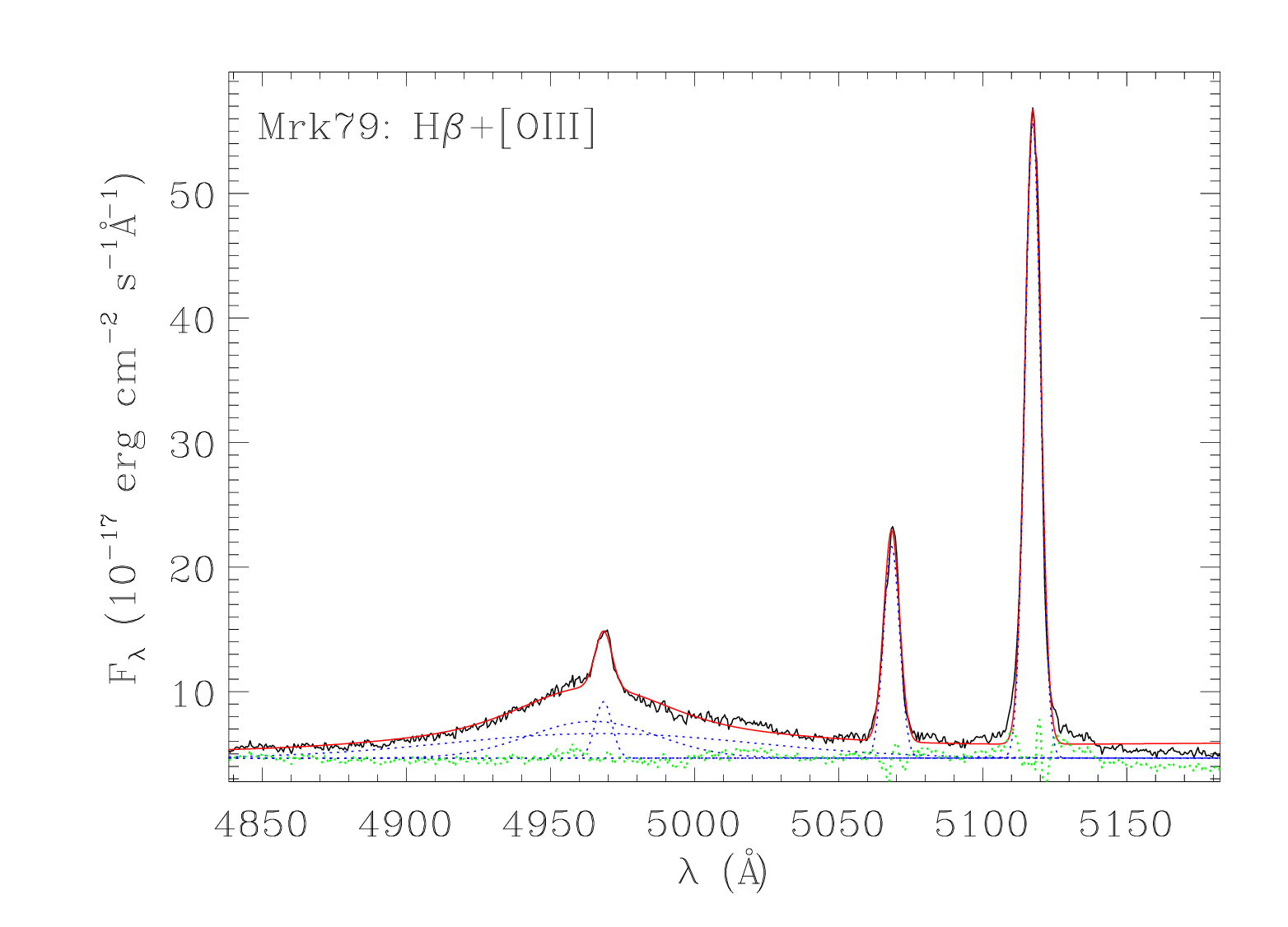}
\includegraphics[scale=0.24]{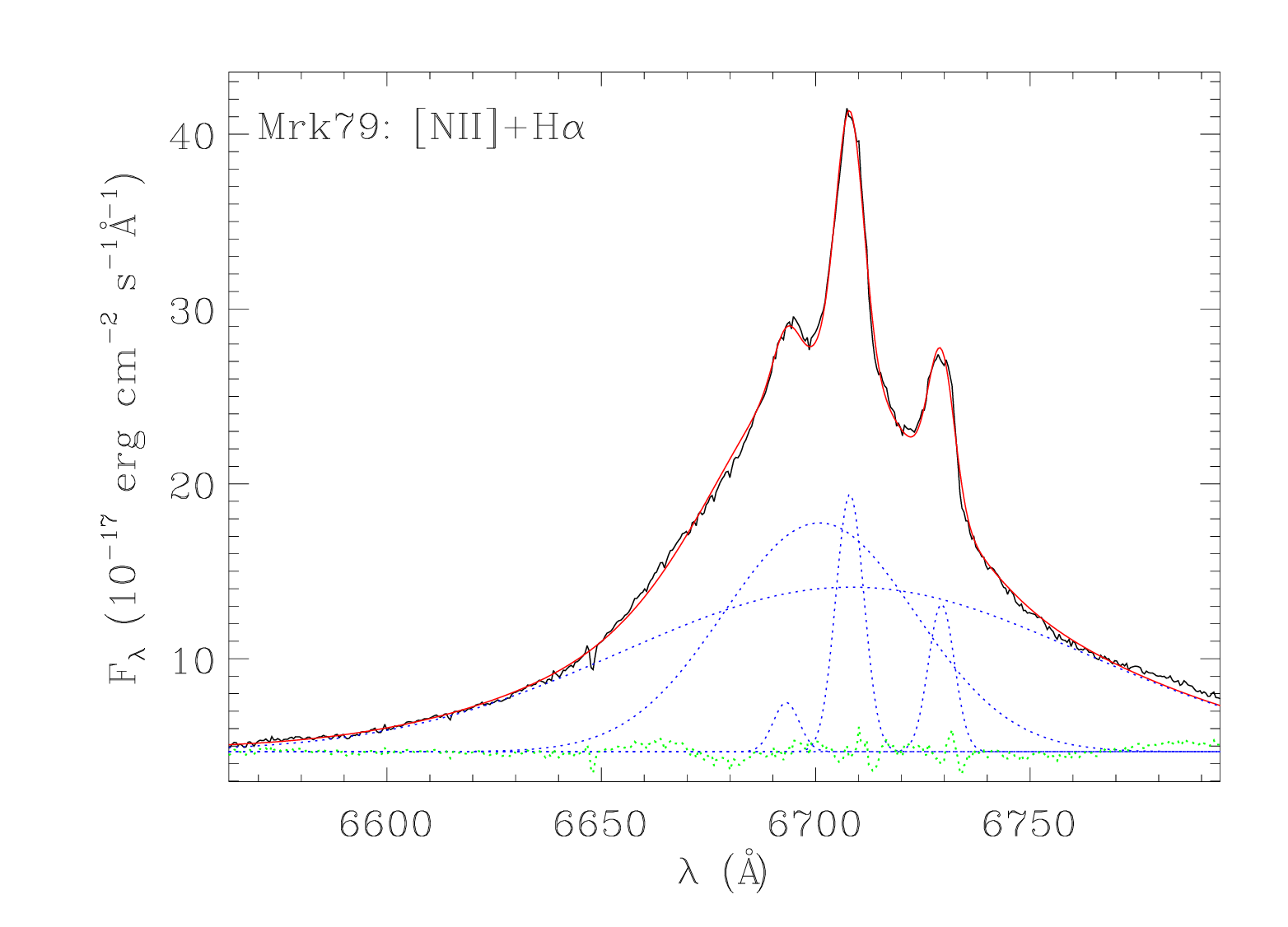}
\includegraphics[scale=0.24]{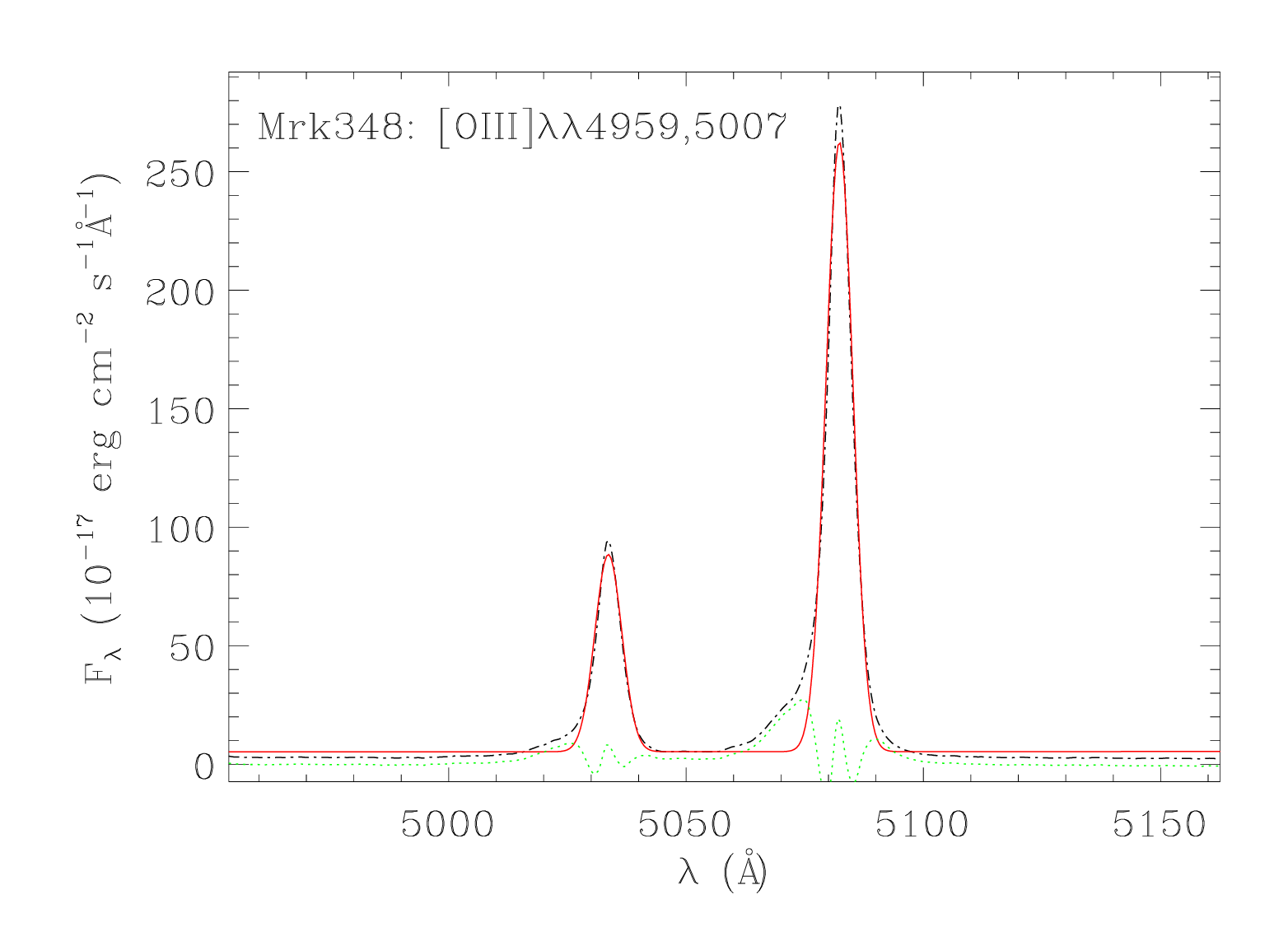}
\includegraphics[scale=0.24]{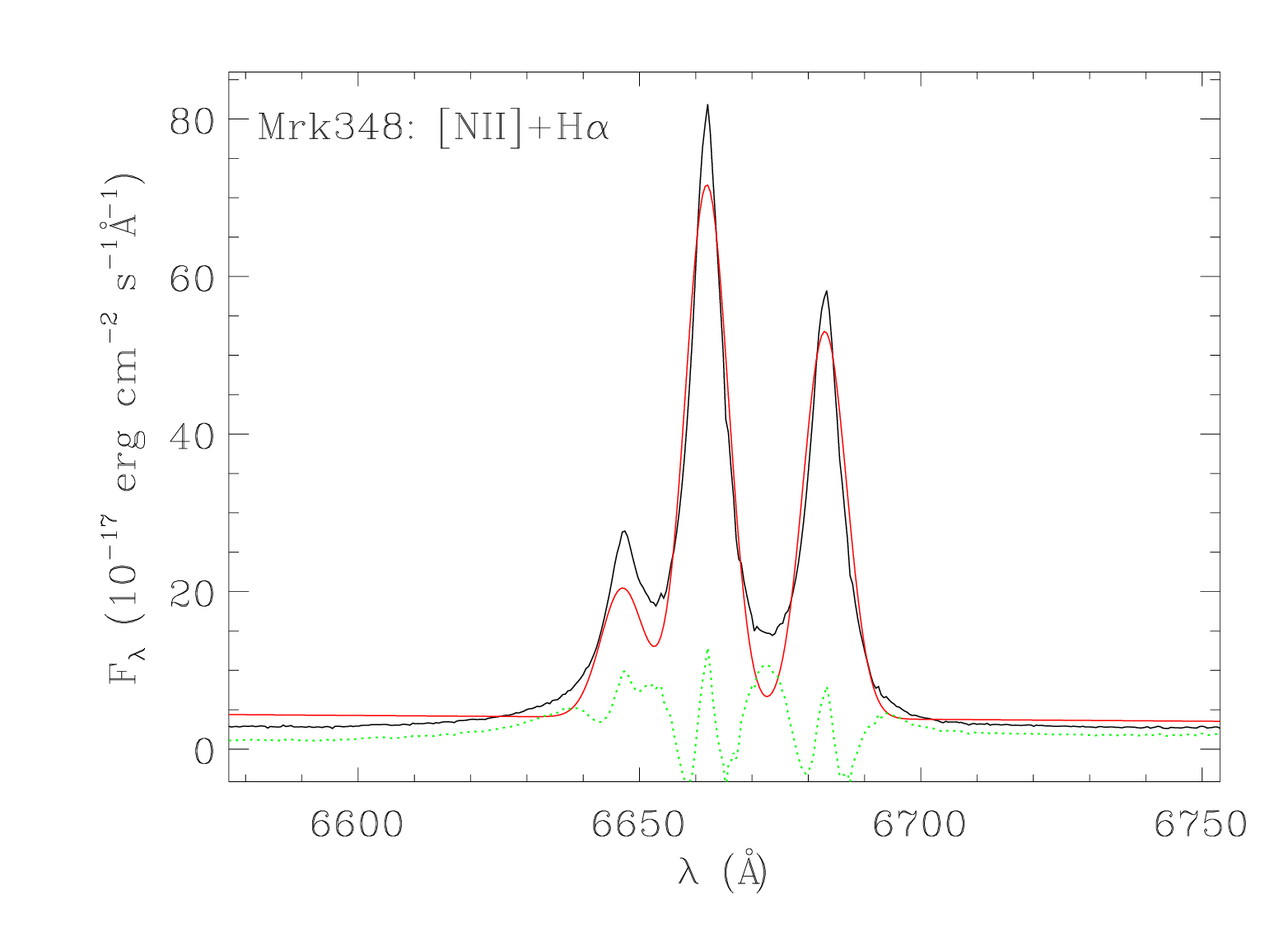}
\includegraphics[scale=0.24]{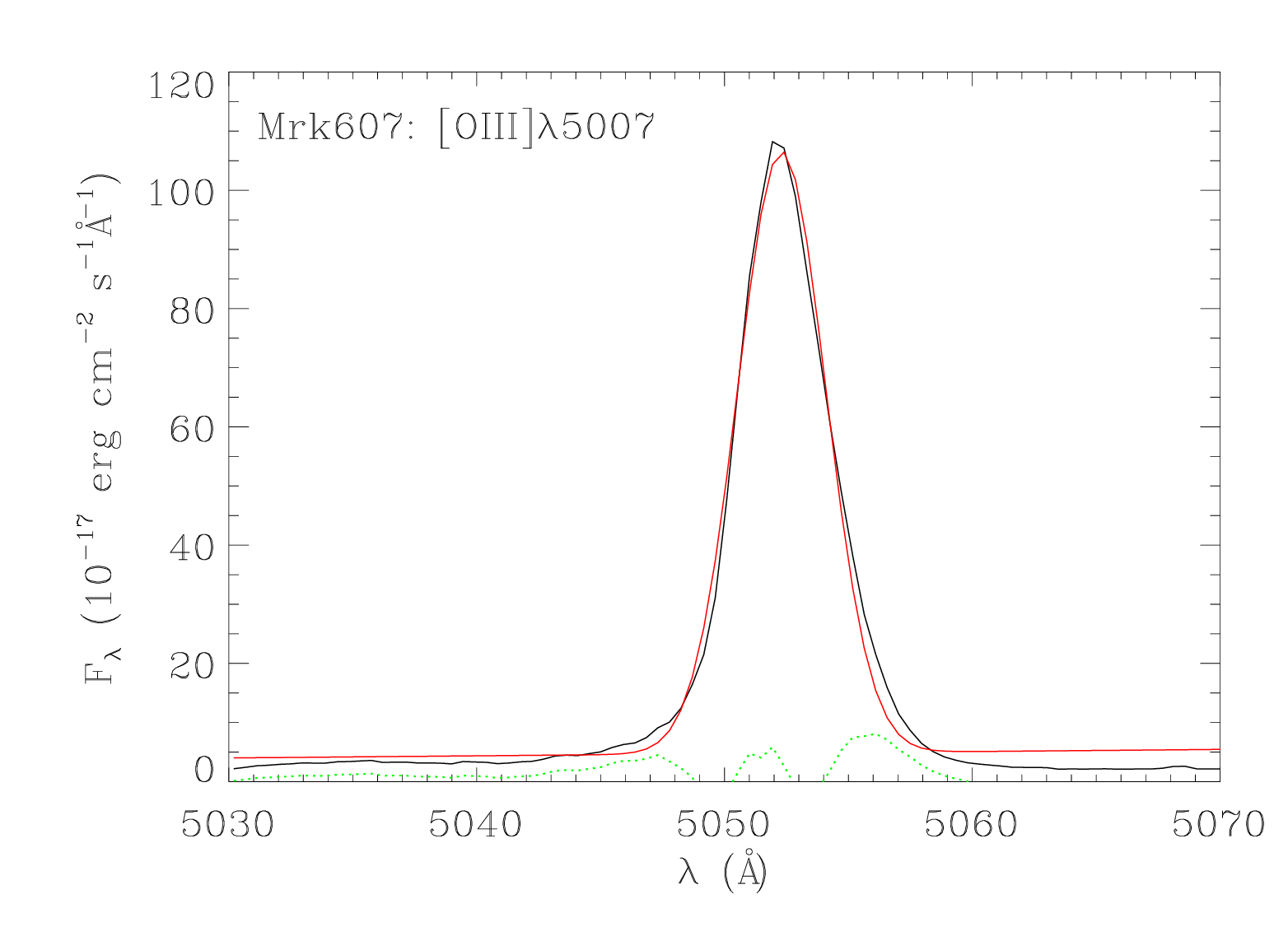}
\includegraphics[scale=0.24]{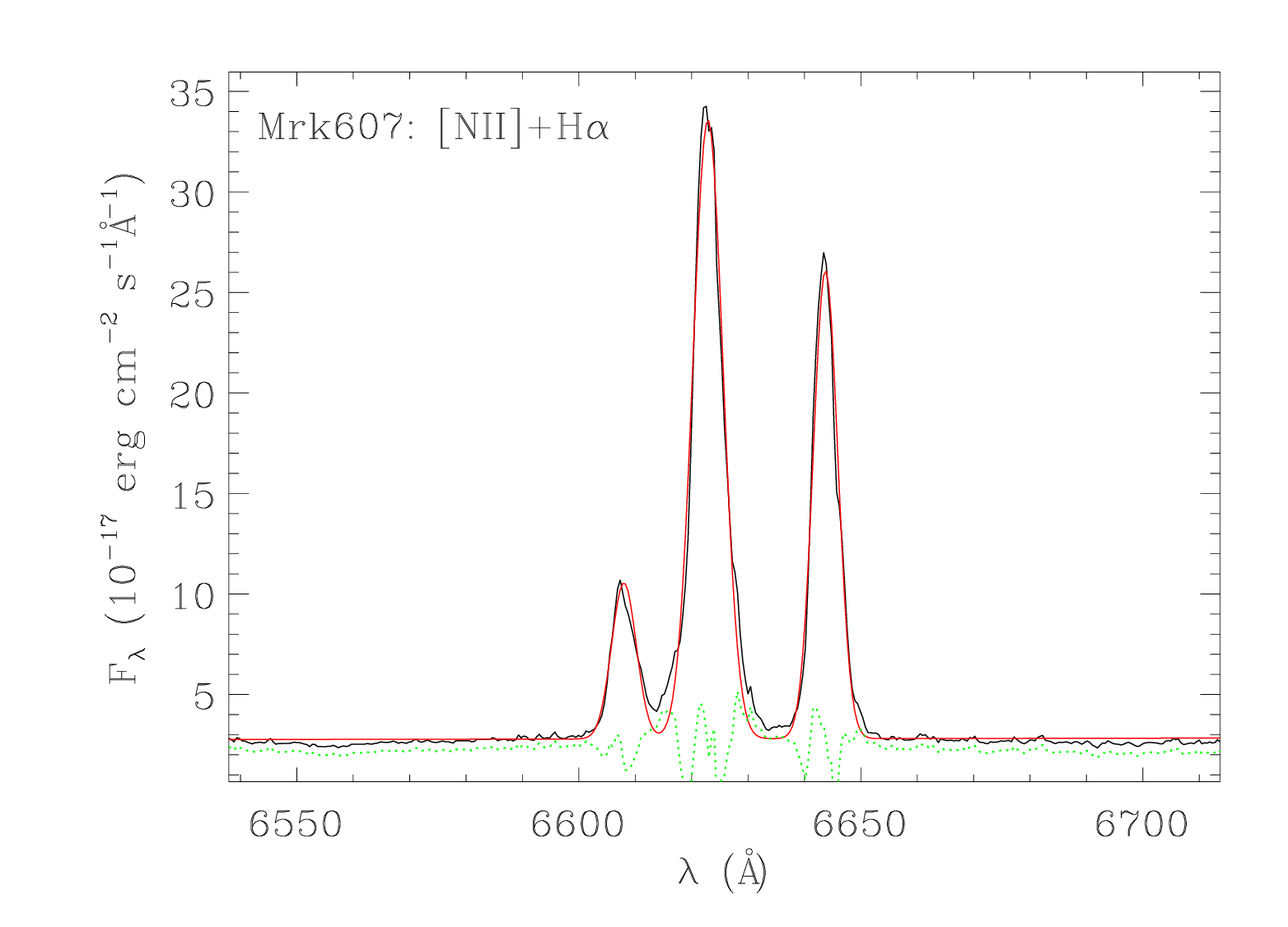}
\includegraphics[scale=0.24]{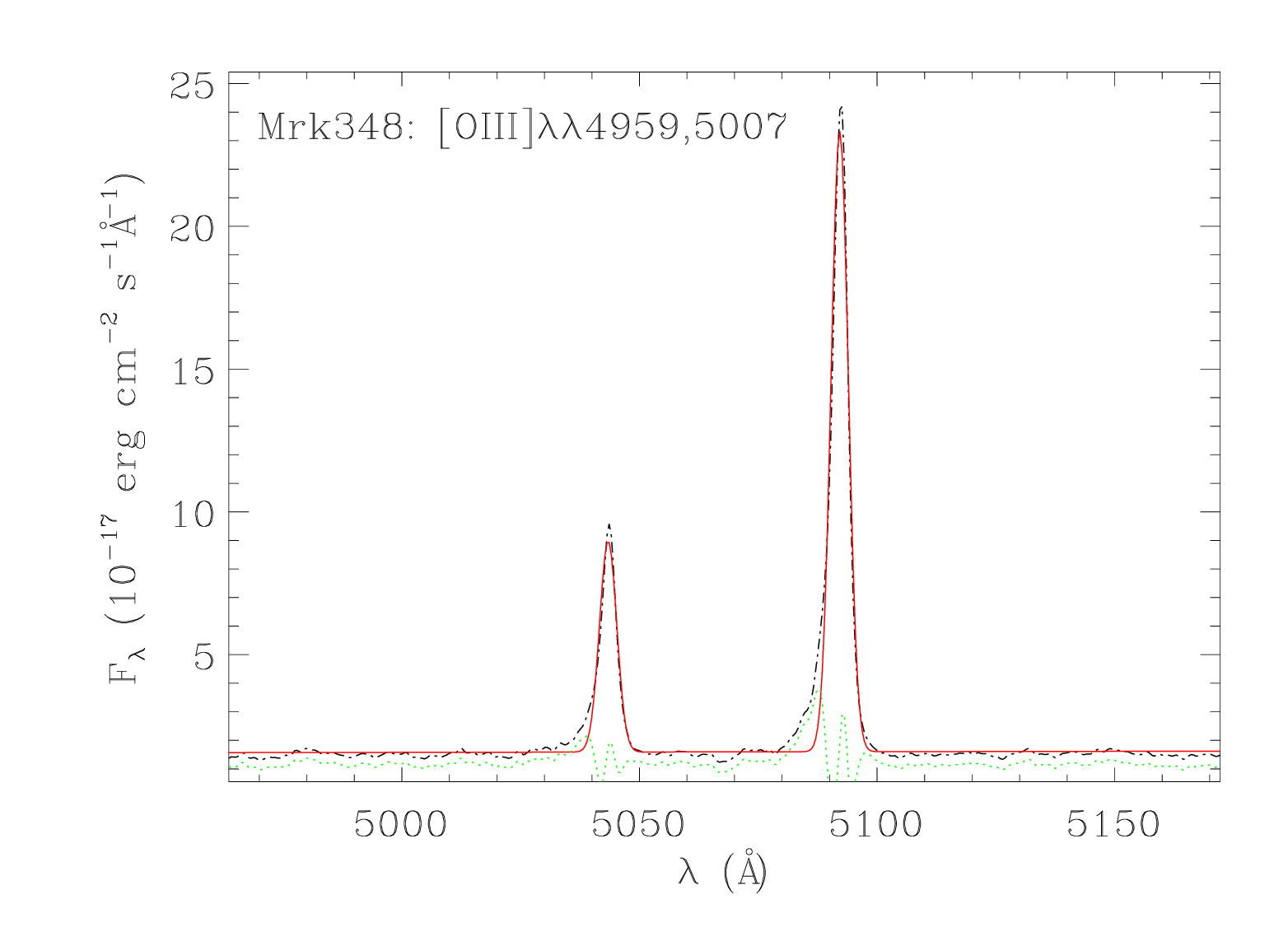}
\includegraphics[scale=0.24]{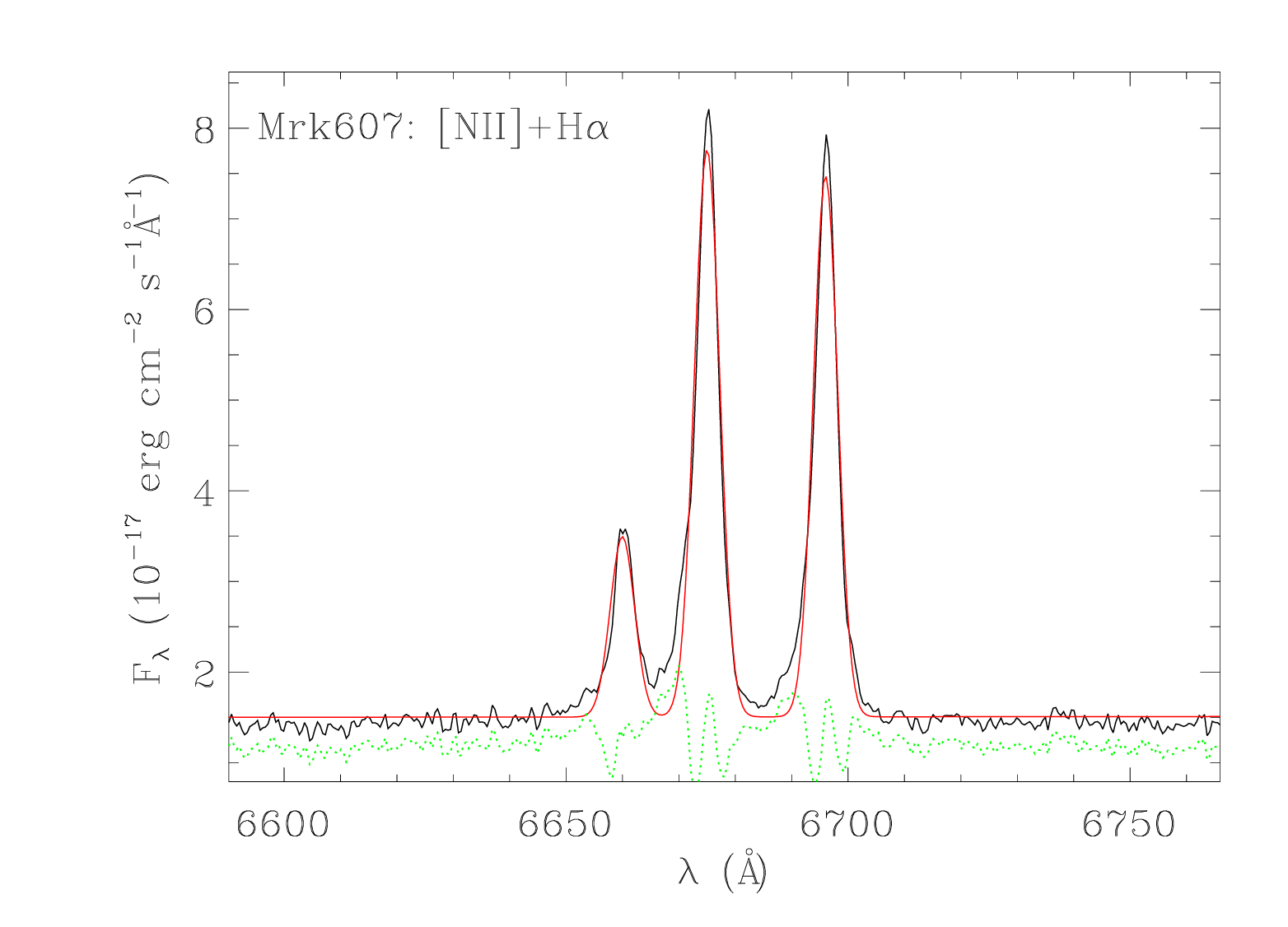}

\caption[spectra]{Examples of fits of the emission line profiles at the nucleus of each galaxy. The galaxy name and fitted emission lines are indicated in the top-left corner of each panel. The observed profiles are shown in black, the best model in red and the dotted green lines show the residuals (observed - model), plus an arbitrary constant. For the Seyfert 1 galaxies (Mrk\,6 and Mrk\,79) each individual component is shown as dotted blue lines. Flux units (F$_{\rm \lambda}$) are 10$^{-17}$ erg\,s$^{-1}$\,cm$^{-2}$\,\AA$^{-1}$.}
\label{fit}
\end{center}
\end{figure*}

In Fig.~\ref{fit} we present 
 examples of the fits of the emission-line profiles for the nuclear spectrum of each galaxy, identified at the top-left corner of each panel. 
 In the case of the Seyfert 1 galaxies, as the broad profiles were clearly non-Gaussian, we had to use two or more components to represent these profiles, and these components, together with the narrow lines are shown as doted blue lines in the Fig.~\ref{fit}.

Although several emission-lines are observed for all objects, we constructed maps for the flux distributions and kinematics of the following emission-lines: [O\,{\sc iii}]\,$\lambda$5007, [O\,{\sc i}]\,$\lambda$6300, H$\alpha$, [N\,{\sc ii}]\,$\lambda$6583 and [S\,{\sc ii}]\,$\lambda$6731. These particular lines were chosen because they present the highest signal-to-noise (S/N) ratio among their species. In addition, we fitted also the H$\beta$ profile, as it can be used together with the H$\alpha$ flux to estimate the reddening.

\subsection{Stellar Kinematics}

 We used the penalized Pixel-Fitting {\sc ppxf} routine \citep{cappellari04} to fit the absorption spectra of the galaxies of our sample and obtain measurements for the  two-dimensional stellar line-of-sight velocity distribution (LOSVD). Each galaxy spectrum is fitted by convolving spectral templates with the corresponding LOSVD, represented by Gauss-Hermite series. As spectral templates we used Single Stellar Populations (SSP)  synthetic spectra selected from \citet{bc03}, which have similar spectral resolution to that of the GMOS data.

As the stellar absorption features seen in our sample spectra are usually weak, we have fitted the whole spectral range at each spaxel for all galaxies, by masking regions with strong emission lines. During the fit we used a truncated second-order Gauss-Hermite series to parameterize the LOSVD, thus resulting in measurements for the stellar LOS velocity (V$_*$) and velocity dispersion ($\sigma_*$). In addition we allowed {\sc ppxf} to include seventh-order  additive Legendre polynomials to better fit the continuum shape and  we used the {\it clean} parameter to reject all spectral pixels deviating more than 3$\sigma$ from the best fit, in order to exclude spurious features and possible weak emission lines.

As output, the {\sc ppxf} code returns measurements of the V$_*$ and $\sigma_*$ at each spatial position and their corresponding uncertainties. These measurements were used to construct two-dimensional maps, which are presented in the next section together with the emission-line maps.  For Mrk\,6, the S/N ratio of the absorption spectra was not high enough to obtain reliable measurements of the stellar kinematics and thus, we do not present maps for this galaxy.

\section{Results}\label{results}

\begin{figure*}
\begin{center}
    \includegraphics[width=\textwidth]{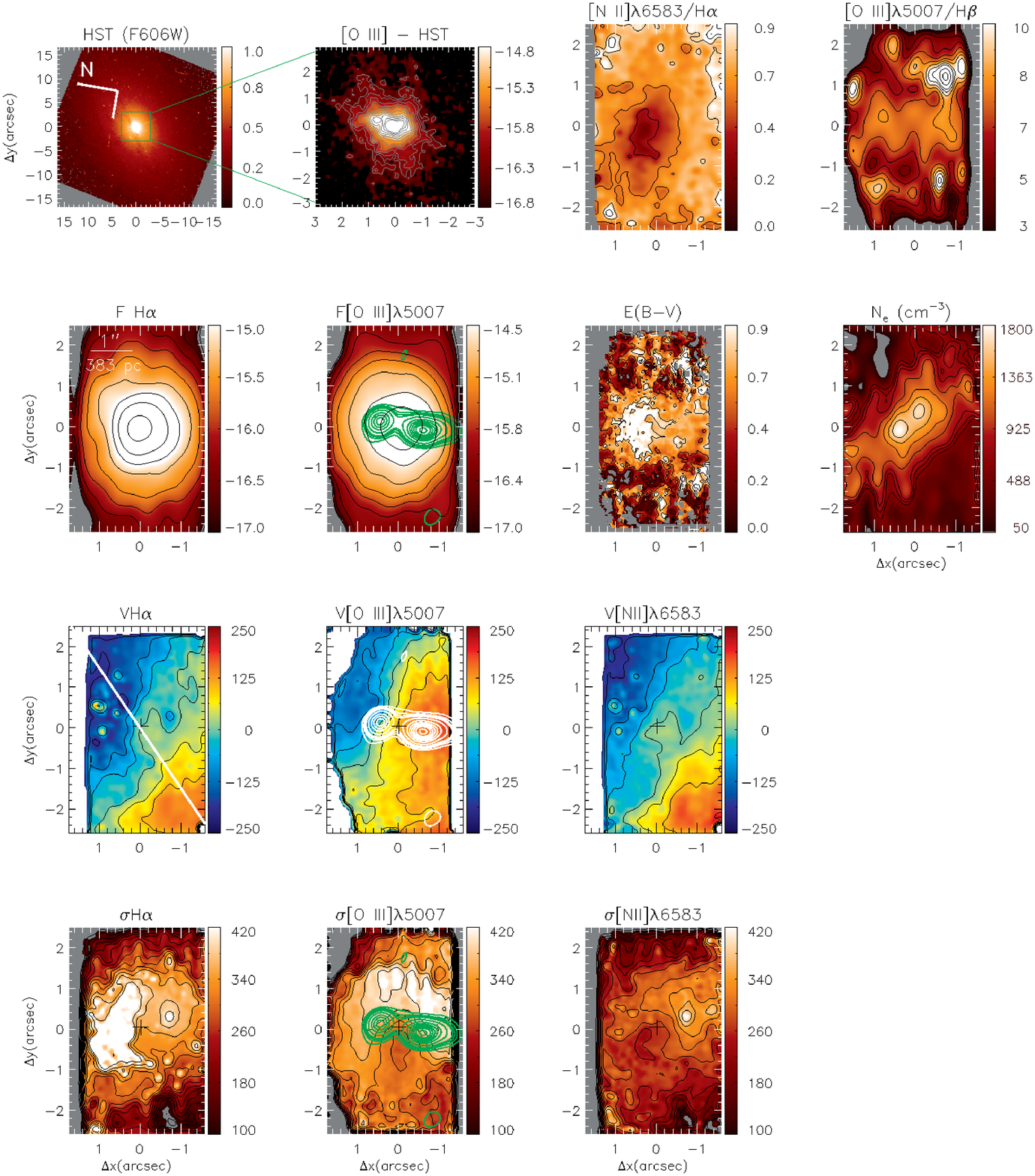}
\centering
\caption{Two-dimensional maps for Mrk\,6 organized as described in Sec.~\ref{results}.
 The central cross marks the location of the nucleus, defined as the position of the peak of continuum emission and the spatial orientation is shown at the top-left corner of the large-scale image. The contours over-plotted to the [O\,{\sc iii}]\,$\lambda$5007 flux map (in green), velocity field (in white) and $\sigma$ map (green) are from the 3.6 cm radio image of \citet{schmitt01}. The white line shown in the H$\alpha$  velocity field represents the major axis of the large-scale disk, measured using I-band images by \citet{schmitt00}. Gray regions in the flux, ratio and velocity dispersion maps and white regions in the velocity fields correspond to locations where the S/N ratio was not high enough to obtain a good fit of the line profiles. For this galaxy we were not able to obtain reliable measurements of the stellar kinematics.}
\label{Mrk6}
\end{center}
\end{figure*}

\begin{figure*}
\begin{center}
    \includegraphics[width=\textwidth]{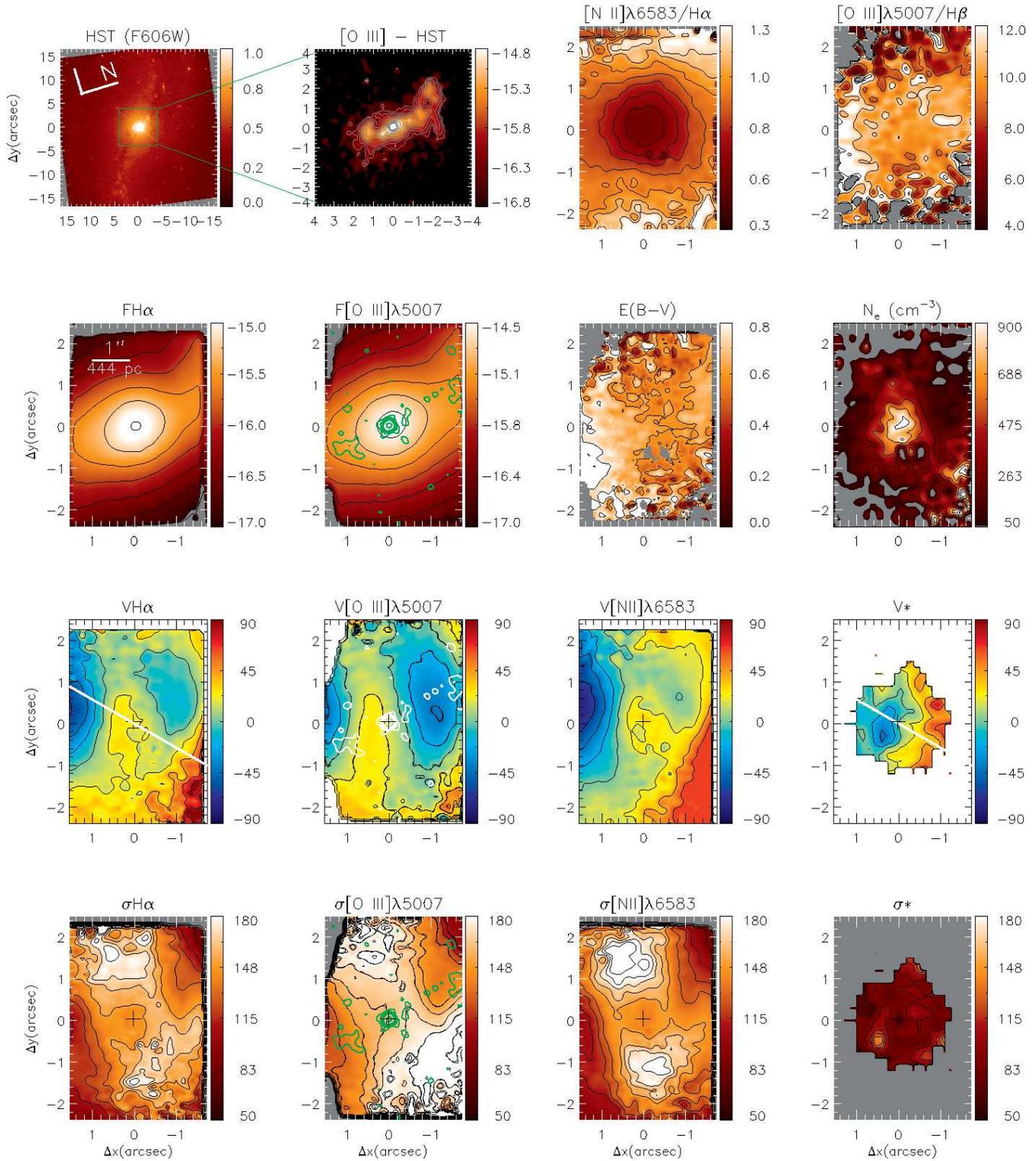}
\centering
\caption{Same as Fig.~\ref{Mrk6} for Mrk\,79. For this galaxy we could obtain reliable measurements for the stellar kinematics within the inner 1$^{\prime\prime}$, and the corresponding stellar velocity field ($V_*$) and velocity dispersion ($\sigma_*$) are shown in the rightmost bottom panels.}
\label{Mrk79}
\end{center}
\end{figure*}

\begin{figure*}
\begin{center}
    \includegraphics[width=\textwidth]{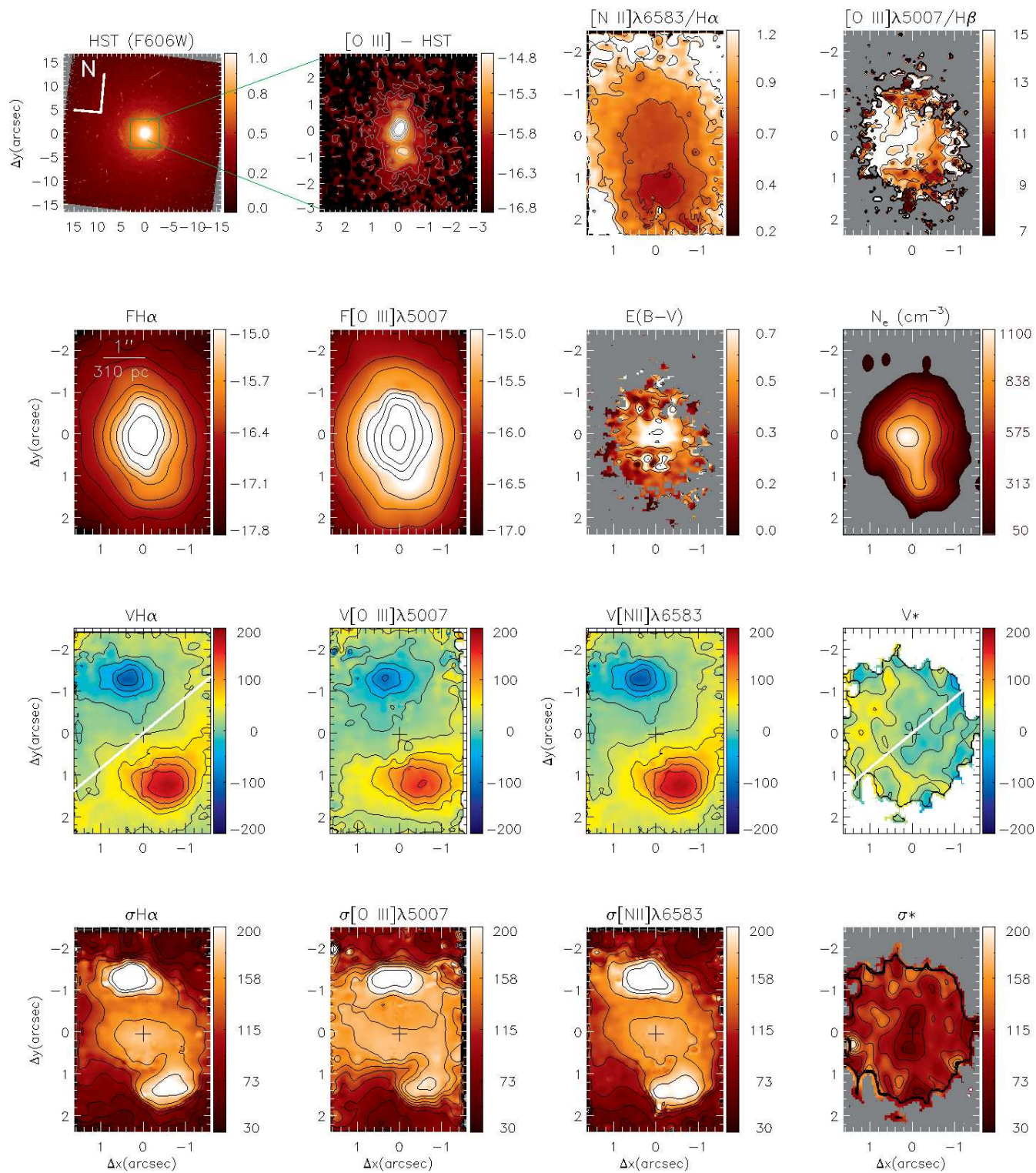}
\centering
\caption{Same as Fig.~\ref{Mrk79} for Mrk\,348.}
\label{Mrk348}
\end{center}
\end{figure*}

\begin{figure*}
\begin{center}
    \includegraphics[width=\textwidth]{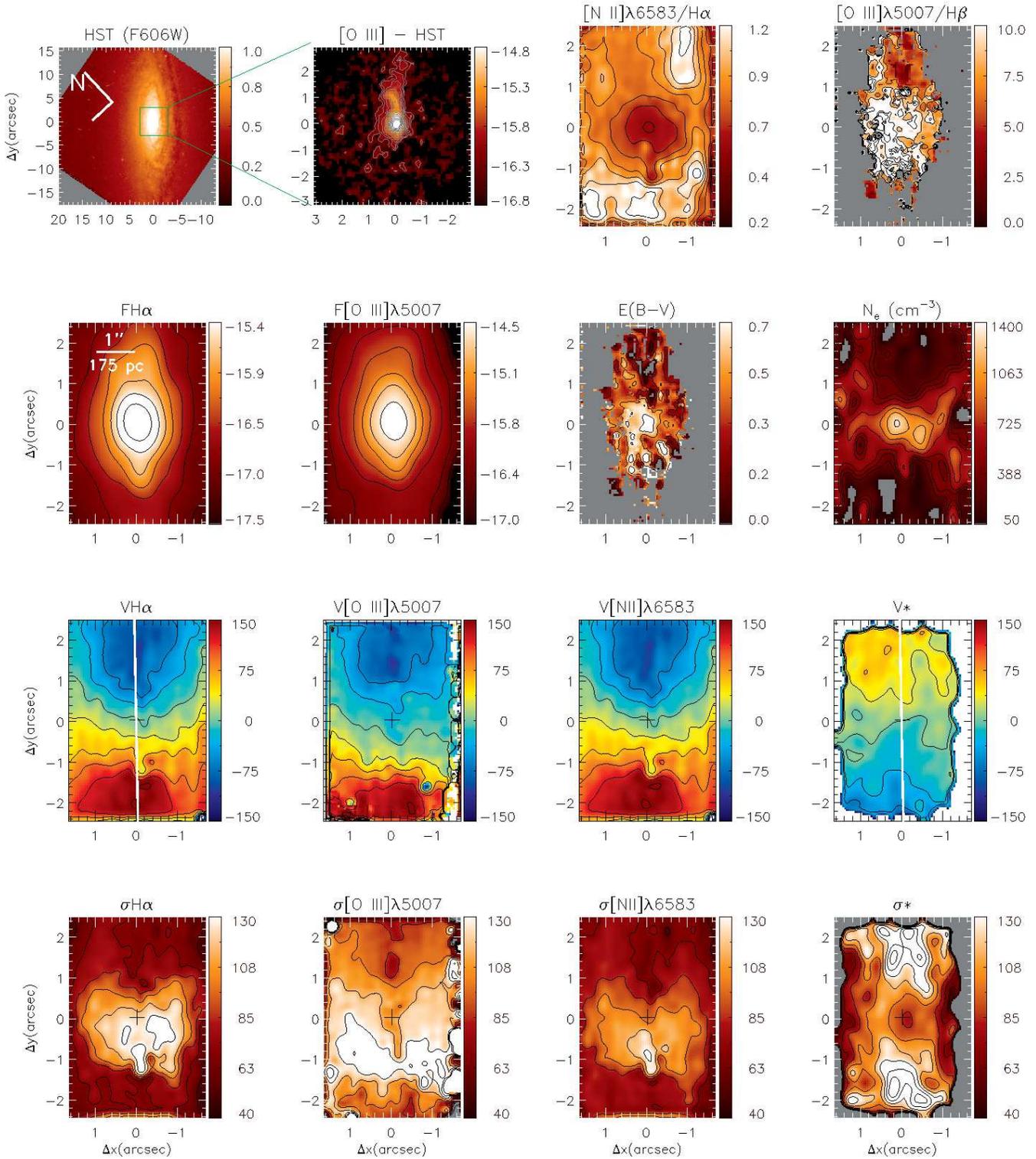}
\centering
\caption{Same as Fig.~\ref{Mrk79} for Mrk\,607.}
\label{Mrk607}
\end{center}
\end{figure*}

\begin{figure*}
\begin{center}
    \includegraphics[width=\textwidth]{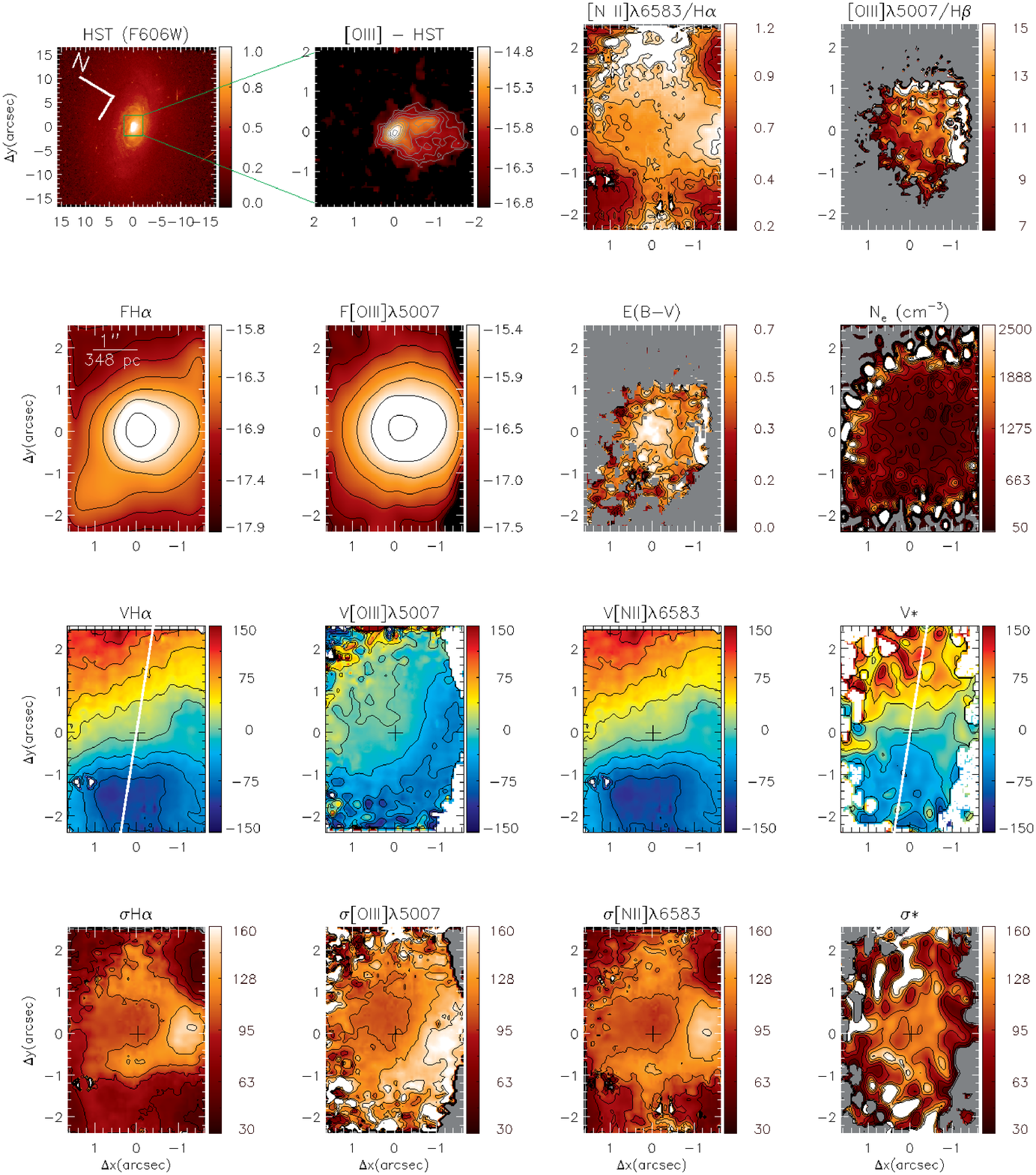}
\centering
\caption{Same as Fig.~\ref{Mrk79} for Mrk\,1058.}
\label{Mrk1058}
\end{center}
\end{figure*}

In Figures~\ref{Mrk6} to~\ref{Mrk1058} we present two-dimensional maps for the flux distributions, gas and stellar kinematics and line ratios for the galaxies of our sample. Each figure is organized as follows:
\begin{itemize}
    \item Top-left panel: large scale image,  obtained with the HST Wide Field Planetary Camera 2 (WFPC2) through the filter F606W \citep{malkan98}. These images are used to illustrate the large scale continuum emission of each galaxy. The spatial orientation of our GMOS data is shown in the top-left corner. The color bar shows the flux in arbitrary units;

    \item  Top-2nd panel (from left to right): \oiii$\,\lambda5007$\,\AA~narrow-band image, obtained with the WFPC2 Linear Ramp Filter \citep[from ][]{schmitt03}. The color bar shows the flux in units of 10$^{-16}{\,\rm erg\,s^{-1}cm^{-2}}$. 
    The images cover the inner 6$^{\prime\prime}\times$6$^{\prime\prime}$ for the galaxies Mrk\,6, Mrk\,348 and Mrk\,607, the inner 8$^{\prime\prime}\times$8$^{\prime\prime}$ for Mrk\,79 and 4$^{\prime\prime}\times$4$^{\prime\prime}$ for Mrk\,1058; the angular resolution is 0\farcs1, about 5 times better than that of our GMOS data;

    \item Top-3rd panel: [N\,{\sc ii}]$\lambda6583$/H$\alpha$ line ratio map;

    \item Top-rigth panel: [O\,{\sc iii}]$\lambda5007$/H$\beta$ line ratio map;

  \item 2nd row, left panel: flux distribution for the H$\alpha$ emission line, obtained from the fit of the line profile with a Gaussian curve. The color bar shows the flux in logarithmic units of ${\rm erg\,s^{-1} cm^{-2}}$; 

  \item 2nd row, 2nd panel: flux distribution for the [O\,{\sc iii}]$\lambda5007$ emission-line, shown in logarithmic units of ${\rm erg\,s^{-1} cm^{-2}}$;

 \item 2nd row, 3rd panel: reddening map obtained from the H$\alpha$/H$\beta$ line ratio using the following expression:

\begin{equation}
E(B-V) = 1.38 \log\left(\frac{\frac{\rm F_{H_\alpha}}{\rm F_{H_\beta}}}{2.86}\right)
\end{equation}
where F$_{\rm H\alpha}$ and F$_{\rm H\beta}$ are the observed fluxes of the H$\alpha$ and H$\beta$ emission-lines, respectively. 
This expression was obtained considering the reddening law of \citet{Cardelli89} and adopting the theoretical line ratio $\frac{\rm F_{H\alpha}}{\rm F_{H\beta}}=2.86$,  corresponding to the case B of H\,{\sc I} recombination for an electron density of $N_e=100$~cm$^{-3}$ and electron temperature of $T_e=10\,000$~K \citep{Osterbrock06};

\item 2nd row, rigth panel: electron density map obtained from the  [S\,{\sc ii}]\,$\lambda\lambda$6716/6731 flux ratio assuming an electron temperature $T_e=10\,000$~K using the $temden$ {\sc iraf} routine.

\item 3rd row, left panel: velocity field for the H$\alpha$ emitting gas (narrow component). The color bar shows the $V_{LOS}$ in units of km\,s$^{-1}$, after the subtraction of the systemic velocity of the galaxy, defined as the velocity measured for the stars from the inner 0\farcs45$\times$0\farcs45, with exception of Mrk~6, for which the sistemic velocity is defined as the velocity of the H$\beta$ narrow component for the same aperture;
\item 3rd row, 2nd panel: same as previous panel, for the [O\,{\sc iii}]$\lambda5007$ emission line;

\item 3rd row, 3nd panel: same as previous panel, for the [N\,{\sc ii}]$\lambda6583$ emission line;

\item 3rd row, rigth panel: same as previous panel, for the stars;

\item Bottom-left panel: velocity dispersion map for the H$\alpha$ emission-line, with the color bar showing the $\sigma$ values in units of km\,s$^{-1}$;

\item Bottom-2nd panel: same as previous panel, for the [O\,{\sc iii}]$\lambda5007$ emission-line;

\item Bottom-3rd panel: same as previous panel, for the [N\,{\sc ii}]$\lambda6583$ emission-line;

\item Bottom-rigth panel: same as previous panel, for the stars.

\end{itemize}

In all maps, the white/gray regions correspond to masked regions where the S/N ratio was not high enough to allow good fits of the emission-line profiles. At other locations, the uncertainties in flux are smaller than 30\,\%  and in velocity/$\sigma$ smaller than 30\,km\,s$^{-1}$.  For most locations the uncertainty in the [O\,{\sc iii}]$\lambda5007$ and H$\alpha$ flux and velocities are smaller than 10\,\% and 15~\,km\,s$^{-1}$, respectively.

The emission-line flux distributions for other emission-lines, nominally  H$\beta$, [O\,{\sc i}]\,$\lambda$6300, [N\,{\sc ii}]\,$\lambda$6583 and [S\,{\sc ii}]\,$\lambda$6731, are shown in Figure~\ref{flux_apen} of Appendix~\ref{apen}. We do not show the kinematic maps for these lines, as they are similar to that of H$\alpha$ and [O\,{\sc iii}]$\lambda5007$ shown in  Figs.~\ref{Mrk6} --~\ref{Mrk1058}.

\subsection{Emission-line Flux Distributions}

All galaxies show extended emission up to the borders of the observed field-of-view (FoV) for the strongest emission-lines, corresponding to typical extents of 0.6 kpc to 1.5 kpc at the galaxies. The line-emission peak is observed at the location of the nucleus for all lines and all galaxies. Mrk\,6 shows an elongated structure to 1\farcs5 (575 pc) north of the nucleus at intermediate flux levels, being more evident for the strongest emission-lines (H$\alpha$, [O\,{\sc iii}]$\lambda5007$ and [N\,{\sc ii}]$\lambda6583$), while at low flux levels the emission is more elongated along the northwest--southeast direction, approximately coincident with the orientation of the major axis of the galaxy as seen in the large-scale image. The elongation to the north is in agreement with that observed in the HST [O\,{\sc iii}] image, although this latter shows several other sub-structures due to its higher spatial resolution.

For Mrk\,79, all lines show similar flux distributions. The flux maps show a curved elongation up to the border of the FoV 1110 pc to the north of the nucleus and a similar elongation to the south. This structure makes an angle of $\approx$ 50$^\circ$ with the large-scale disk major axis, as seen in the HST F606W image. This structure is observed in the HST narrow-band image as a much more collimated and with smaller scale features due to the better spatial resolution.

The H$\beta$ and [O\,{\sc i}]\,$\lambda$6300\, emission for Mrk\,348 (Fig.~\ref{flux_apen}) are restricted to the inner $\approx$ 1\farcs5 (465 pc), while others maps show extended emission over the whole FoV (1.1 kpc x 1.5 kpc), with the highest levels being more elongated along the northeast-southwest direction. The [O\,{\sc iii}] HST image shows two blobs of higher emission, one centred at the nucleus and another at $\approx$ 1\farcs0 southwest of it.

Mrk 607 shows extended emission  over the whole FoV (0.6 kpc x 0.9 kpc) for most emission-lines. The highest flux levels are more extended along the southeast-northwest direction, in agreement with that observed in the HST [O\,{\sc iii}] image. The H$\beta$ and [O\,{\sc i}]\,$\lambda$6300\, emission maps are more compact, with the [O\,{\sc i}] one being showing a north-south elongation within the inner $\approx$ 0\farcs8 (140 pc).

For Mrk\,1058, the highest flux levels for the [O\,{\sc iii}] emission are more elongated to the southwest, in agreement with the HST [O\,{\sc iii}] image, that shows a structure there at 1\farcs0 (350\,pc)  from the nucleus. The other flux distributions (e.g. H$\alpha$) show also extended emission approximately along the west-east direction up to the borders of the FoV.

\subsection{Line-of-sight Velocity Maps}\label{VLOS}

The $V_{LOS}$ maps for all galaxies are shown in Figures~\ref{Mrk6} -- \ref{Mrk1058}. The velocities are shown relative to the systemic velocity of each galaxy, adopted as the value obtained for the stellar velocity within the inner 0\farcs45$\times$0\farcs45, with exception of Mrk~6, for which the adopted value corresponds to the central wavelength of the narrow component of the H$\beta$ emission-line of the nuclear spectra for the same aperture. The derived heliocentric systemic velocities are 5603\,\kms, 6545\,\kms{}, 4473\,\kms, 2782\,\kms{} and 5053\,\kms{} for Mrk\,6, Mrk\,79, Mrk\,348, Mrk\,607 and Mrk\,1058, respectively. 

The $V_{\rm LOS}$ maps for Mrk\,6 (Fig.~\ref{Mrk6}) present a distorted rotation pattern with blueshifts observed to the north of the nucleus and redshifts to the south of it, with a projected velocity amplitude of $\approx$ 200 km\,s$^{-1}$. The distortions in the rotation pattern differ between the H$\alpha$ and [O\,{\sc iii}] rotation fields: while the highest redshifts are observed at $\approx$  1\farcs0 south of the nucleus for the [O\,{\sc iii}], for H$\alpha$ the highest velocities are observed next to the corner of the IFU FoV at $\approx$  2\farcs0 southeast of the nucleus. The [N\,{\sc ii}] velocity field is similar to that of [O\,{\sc iii}] and we were not able to map the stellar kinematics for this galaxy, due to the low S/N ratio of the stellar absorption features. 

The most conspicuous structures observed in the Mrk\,79 gas velocity fields are two ``blobs'' observed in blueshifts, of up to $-90$ km\,s$^{-1}$, one at $\approx$ 1\farcs0 (444 pc)  to the north and another at $\approx$ 1\farcs5 (666 pc) to the south-southeast of the nucleus, at locations coincident with blobs seem in the HST [O\,{\sc iii}] image. In addition, some redshifts are observed to west of the nucleus at distances larger than 2\farcs0 from it, more clearly seen in the H$\alpha$ $V_{\rm LOS}$ map. We were able to measure the stellar kinematics only within the inner 1$^{\prime\prime}$ and the corresponding stellar $V_{LOS}$ map show blueshifts to the southeast and redshifts to the northwest, approximately along the orientation of the photometric major axis of the galaxy.

In the case of Mrk\,348 (Fig.~\ref{Mrk348}), the main gas kinematics structures are  two blobs observed at $\approx$ 1\farcs5 (465 pc) to the northeast and southwest of the nucleus, observed in all $V_{LOS}$ maps. The blob to the northeast shows blueshifts of up to -200 km\,s$^{-1}$ while the one to the southwest shows redshifts with similar amplitudes. The emission-line profiles at these locations are very complex and not well reproduced by a single gaussian component, possible indicating the presence of gas outflows from the galaxy nucleus and suggest a bipolar outflow from the nucleus.  The stellar velocity field shows a smaller velocity amplitude of $\sim$50\,\kms\ with redshifts observed to the east and blueshifts to the west.

A clear rotation pattern is observed for the gas in Mrk\,607 (Fig.~\ref{Mrk607}), with the line of nodes oriented along the PA $\approx$ $-60/120^\circ$, with blueshifts observed to the northwest and redshifts to the southeast. The observed gas velocity amplitude is $\approx$ 150 km\,s$^{-1}$. A rotation pattern is also observed for the stars, but at the opposite orientation, with blueshifts seen to the southeast and redshifts to the northwest. In addition, the stellar $V_{LOS}$ map shows a smaller velocity amplitude of $\sim70$\,\kms.

The stellar $V_{LOS}$ map for Mrk\,1058 shows a rotation pattern along the photometric major axis of the galaxy, with blueshifts seen to the  the southeast and redshift to the northwest, with a projected velocity amplitude of $\sim$120\,\kms. We see a distorted rotation pattern in the H$\alpha$ and [N\,{\sc ii}] velocity fields with blueshifts and a steeper rotation pattern to the southeast and redshift to the northwest, and a projected velocity amplitude of $\approx$ 150 km\,s$^{-1}$. In the case of [O\,{\sc iii}], the velocity field seems more disturbed, with blueshifts dominating the velocities. In particular, blueshifts observed at $\approx$ 1\farcs5 (522 pc) southwest of the nucleus (less conspicuous but also present in the H$\alpha$ velocity field) are co-spatial with the elongation seen in the [O\,{\sc iii}] images, suggesting it is an outflow from the nucleus.

\subsection{Velocity Dispersion Maps}\label{sigma}

The gas and stellar velocity dispersion maps for Mrk\,6, Mrk\,79, Mrk\,348, Mrk\,607 and Mrk\,1058 are shown in the bottom panels of Figures~\ref{Mrk6} --~\ref{Mrk1058}.

Mrk\,6 (Fig.~\ref{Mrk6}) shows $\sigma$ values ranging from $\approx$ 100 km\,s$^{-1}$ to $\approx$ 450 km\,s$^{-1}$. The highest $\sigma$ values for H$\alpha$ are seen in a broad arc-shaped region centered at 0\farcs7 to the north of the nucleus (bottom-left panel of Fig.~\ref{Mrk6}). This structure seems to correspond to the extended emission observed in the HST \oiii\ image. The \oiii\ $\sigma$ map also shows an arc-shaped structure, partially co-spatial with that observed in H$\alpha$, but the highest values are observed (at similar distances) to the west of the nucleus. The [N\,{\sc ii}] shows overall smaller values of $\sigma$ than [O\,{\sc iii}], but with the highest values observed at the same position to the west.

The gas $\sigma$ maps for Mrk\,79 (Fig.~\ref{Mrk79}) show values from 80 to 180\,\kms, with the \oiii\ showing systematically larger values than H$\alpha$ and [N\,{\sc ii}]. This may indicate that the [O\,{\sc iii}] emitting gas is located at larger latitudes, while the H$\alpha$ and [N\,{\sc ii}] emission have a more important contribution from gas located close to the plane of the disk.

The highest values are observed in a bipolar structure oriented along the east-west direction, perpendicular to the direction of the strongest  [O\,{\sc iii}] and H$\alpha$ emission. The stars show small $\sigma$ values at all locations, with $\sigma < 100$\,\kms.  

The $\sigma$ values for the emission lines of  Mrk~348 (Fig.~\ref{Mrk348}) are in the range: 50 -- 200~\kms, with the highest values observed at the blueshifted and redshifted blobs seen in the velocity fields at $\approx$ 1\farcs5 NE and $\approx$ 1\farcs3 southwest of the nucleus, respectively. This is consistent with the interpretation that they correspond to outflowing gas. At locations closer to the nucleus, the $\sigma$ values are between 150 and 200 km\,s$^{-1}$, while the lowest values are observed in regions beyond the blobs. The stellar velocity dispersion map presents values smaller than 150~\kms\ at all locations, showing a drop at the nucleus with $\sigma\sim50$\,\kms.        

Mrk\,607 (Fig.~\ref{Mrk607}) shows the lowest average gas $\sigma$ values of our sample, ranging from 50 to 130~\kms. The highest $\sigma$ values are observed in a strip perpendicular to the galaxy major axis, between the nucleus and $\approx$ 1\farcs3 (227 pc) to the southwest, opposite (relative to the nucleus) to the main elongation seen in the \oiii\ flux map. A distinct behaviour is seen at the $\sigma_*$ map, that presents the highest values at distances of 1--2$^{\prime\prime}$ to the northwest and to the southeast, along the major axis of the galaxy.

For Mrk~1058 (Fig.~\ref{Mrk1058}), H$\alpha$ and [N\,{\sc ii}] $\sigma$ maps shows somewhat smaller values than the \oiii\ map (40 -- 120 km\,s$^{-1}$ vs. 40 -- 150 km\,s$^{-1}$). The highest values are observed $\approx$ 1\farcs3 (452 pc) to the southwest of the nucleus, at the location where excess blueshifts and extended \oiii\ emission is observed, consistent with the presence of an outflow. A partial ring of intermediate $\sigma$ values seems to surround the nucleus at 1\farcs0.  This partial ring is not observed in the stellar $\sigma$ map, which shows values smaller than 130~\kms\ at most locations.

\subsection{Line Ratio Maps and Diagnostic Diagram}

\begin{figure}
\begin{center}
    \includegraphics[scale=0.5]{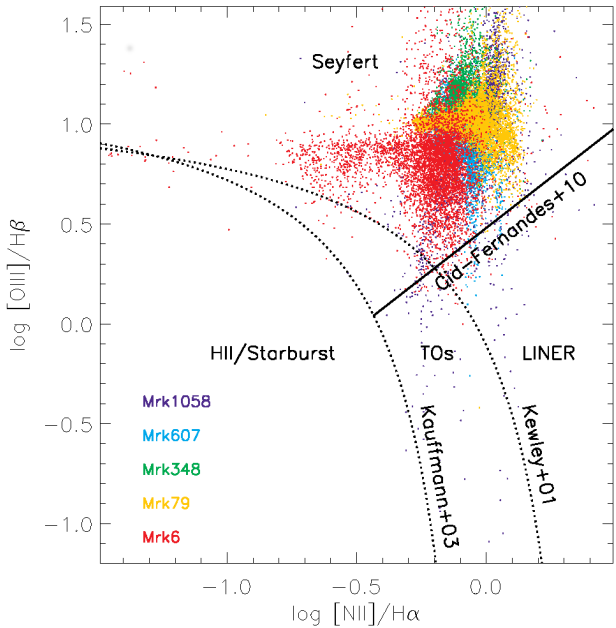}\centering
\caption[BPT Diagram]{\small BPT diagram [O\,{\sc iii}]\,$\lambda$5007/H$\beta$ versus [N\,{\sc ii}]\,$\lambda$6583/H$\alpha$ line ratios for all the observed galaxies. Each galaxy is represented by a colour.}
\label{bpt}
\end{center}
\end{figure}

The emission-line ratios [N\,{\sc ii}]\,$\lambda$6583/H$\alpha$ and [O\,{\sc iii}]\,$\lambda$5007/H$\beta$ are frequently  used to map the NLR excitation via the BPT diagram \citep{Baldwin81}, allowing to distinguish the gas excitation as characteristic of Seyfert, Starburst, LINER or Transition Objects (TOs). 
We have obtained this diagram for the central regions of the galaxies of our sample in Figure~\ref{bpt}, with each spaxel corresponding to a point in the diagram. We included in this plot only regions where the flux uncertainties of the four emission-lines are smaller than 30\,\%, in order to avoid spurious features. The value of these ratios for the spaxels of each galaxy are shown as points with distinct colors:  Mrk\,6  in red, Mrk\,79 in yellow, Mrk\,348  in green, Mrk\,607 in blue and Mrk\,1058 in purple. The dashed curves represent the division lines between star forming galaxies and AGN from \citep{Kauffmann03} and \citep{Kewley01}. The solid line shows the division between Seyfert and LINER nuclei \citep{cf10}.

For the five galaxies, all spaxels are observed in the Seyfert region of the BPT diagram, showing that the gas excitation in the inner kiloparsec of all galaxies of the sample is dominated by  radiation from the AGN.

 Although the BPT diagram shows only points within the Seyfert region for all galaxies, spatial variations of the excitation of the NLR can be observed in the [N\,{\sc ii}]/H$\alpha$ and [O\,{\sc iii}]/H$\beta$ maps of Figs.~\ref{Mrk6} to \ref{Mrk1058}.


For Mrk~6, the [N\,{\sc ii}]/H$\alpha$ map  (Fig.~\ref{Mrk6}) shows values ranging from 0.4 to 0.8, with most values close to the maximum and the lowest values observed mostly to the northeast of the nucleus, approximately co-spatial with the arc region showing the highest $\sigma$ values in H$\alpha$ and also with the highest reddening (see next section). [O\,{\sc iii}]/H$\beta$ shows values around 8 in an $\approx$\,1\farcs0 wide stripe passing through the nucleus from the northeast to the southwest, approximately following the orientation of the HST [O\,{\sc iii}] image. Lower values down to approximately 0.5 are seeing surrounding the strip, with a few spots with higher values. 

 The [N\,{\sc ii}]/H$\alpha$ ratio map for Mrk~79  (Fig.~\ref{Mrk79}) shows the smallest values ($\approx$ 0.3) at the nucleus and  increasing outwards, up to 1.3 at distances larger than 2\farcs0.  The [O\,{\sc iii}]/H$\beta$ map presents values in the range 4 -- 11, with values of about 10 approximately along the north-south direction, following the orientation of the collimated \oiii\ emission observed in the HST narrow-band image, and are surrounded to the east and west by the smallest ratios of the order of 6. The highest values ($\ge$ 12) are observed to the south at the border of the field and distances of $\ge$ 1\farcs0.

Mrk~348 (Fig.~\ref{Mrk348}) shows [N\,{\sc ii}]/H$\alpha$ ratio ranging from 0.6 to 1.2. The smallest values are observed at approximately 1\farcs5 to the southwest, at a location just beyond the extended emission observed in the HST [O\,{\sc iii}] map and the GMOS H$\alpha$ and [O\,{\sc iii}] maps. The highest values are observed to the north-northeast of the nucleus, at distances of $\approx$ 2\farcs0 from the nucleus. Due to the low S/N  ratio of the H$\beta$ emission-line, we were able to obtain the [O\,{\sc iii}]/H$\beta$ only for the inner 1\farcs5. The map has the highest values ($\approx$ 15) to the northeast and the lowest ratios  ($\approx$ 10) to the southwest. 

Fig.~\ref{Mrk607} shows that the  [N\,{\sc ii}]/H$\alpha$ map for Mrk\,607 presents the lowest values (0.5 -- 0.7) at the nucleus and that this value increases outwards to higher than 1.2 in two ``strips" at  $\approx$ 2\farcs0 southeast and $\approx$ 2\farcs0 west of the nucleus. The  [O\,{\sc iii}]/H$\beta$ map shows values ranging from 3 to more than 10, with the highest ones observed at the nucleus and to the north-northeast and east, and the lowest values seen to the northwest.

Mrk\,1058 (Fig.~\ref{Mrk1058})  shows [N\,{\sc ii}]/H$\alpha$ ranging from 0.4 to 1.2, with the smallest values observed to the east, south and west at distances larger than 1\farcs0 from the nucleus, at locations where the H$\alpha$ flux distribution presents elongated structures. The highest values for this ratio are seen at $\approx$ 1\farcs0 southwest and  $\approx$ 2\farcs0 northwest of the nucleus, while intermediate values (0.9 -- 1.0) are observed between these locations. 
 The [O\,{\sc iii}]/H$\beta$ ratio map presents a gradient along the east-west direction (similar to the orientation of the extended H$\alpha$ emission), with lowest values ($\approx$ 9) to the east and the highest values ($\approx$ 15) to the west.

 \subsection{Reddening Maps}
 
 The reddening maps obtained from the \ha/\hb\ line ratio are presented in the top-right panels of Figs.~\ref{Mrk6} -- \ref{Mrk1058}. Mrk\,6 (Fig.~\ref{Mrk6}) presents $E(B-V)$ values ranging from 0.2 to 0.9, with the highest values observed from the nucleus up to 1$\farcs0$ to the northeast, at the location where an increase in the H$\alpha$ velocity dispersion is observed.
 
 For Mrk\,79 (Fig.~\ref{Mrk79}) the $E(B-V)$ values range from 0.4 to 0.8 with the smallest values close to the nucleus and increasing outwards along the direction of extended emission in [O\,{\sc iii}], particularly to the south-southwest at a region of enhanced [O\,{\sc iii}]/H$\beta$ ratio and blueshifts observed in the H$\alpha$ and [O\,{\sc iii}] velocity fields.

 Mrk\,348 (Fig.~\ref{Mrk348}) shows $0.2<E(B-V)<0.7$, with the highest values observed at the nucleus up to $\approx$ 1\farcs0 from it along the east-southeast to west-northwest, and at the location where an extension of the [O\,{\sc iii}] HST emission is observed to the south-southwest ($\approx$ 0\farcs8 from the nucleus).

The $E(B-V)$ map for Mrk\,607 (Fig.~\ref{Mrk607}) shows values in the range 0.2 -- 0.7, with the highest values observed at the nucleus up to $\approx$ 1\farcs0 to the northeast and the lowest values seen to the opposite side and to the northwest, along the major axis of the galaxy.

Mrk\,1058  (Fig.~\ref{Mrk1058}) shows $E(B-V)$ smaller than 0.5 at most locations. The highest values of up to 0.7 are observed at the nucleus and within a radius of $\approx$ 0\farcs4 from it.

\subsection{Electron Density}

We used the  $temden$ {\sc iraf} routine to obtain electron density ($N_e$) maps using the  [S\,{\sc ii}]\,$\lambda\lambda$6716/6731 flux ratio and assuming an electronic temperature $T_e$ = 10\,000\,K. The resulting maps for each galaxy are shown in the bottom-right panel of Figs.~\ref{Mrk6} -- \ref{Mrk1058}.

For Mrk\,6 (Fig.~\ref{Mrk6}), the $N_e$ map shows values ranging from 100 to 1800 cm$^{-3}$, with higher values  observed within $\approx$ 1\farcs0 from the nucleus along the northeast-southwest direction. The highest values are seen in two blobs located at 0\farcs25 northeast and 0\farcs45 southwest of the nucleus. At the nucleus $N_e\approx1300$\,cm$^{-3}$. 

The $N_e$ map for Mrk\,79 (Fig.~\ref{Mrk79}) shows $100<N_e<900$, with the highest values observed within $\approx$ 0\farcs4 from the nucleus. Some high values are also observed outwards towards the east and west, co-spatial regions with enhanced gas velocity dispersion in \oiii$\lambda$5007 and H$\alpha$. 

Mrk\,348 (Fig.~\ref{Mrk348}) shows $N_e$ in the range 100 -- 1100 cm$^{-3}$, with the highest values observed at the nucleus. The $N_e$ map shows an elongated structure to the southwest where $N_e\approx850$\,cm$^{-3}$ at 1\farcs5 SW of the nucleus. This peak is observed co-spatially with an enhancement in the velocity dispersion and a redshifted blob in the gas velocity fields. 

The electron density map for Mrk~607 (Fig.~\ref{Mrk607}) shows values from 100 to up to 1400~cm$^{-3}$ with the highest values at the nucleus. An elongated structure with extent $\approx$ 1\farcs2 and 0\farcs5 width with $N_e\approx$ 1100\,cm$^{-3}$ is observed along the northeast-southwest direction, $\approx$ perpendicularly to the major axis of the galaxy, and along the direction where the highest [O\,{\sc iii}]/H$\beta$ ratios are observed. Enhancements in the velocity dispersions of both [O\,{\sc iii}]  and H$\alpha$ are observed contiguous to this structure to the southeast.

Mrk\,1058 (Fig.~\ref{Mrk1058}) presents the highest values of $N_e$ among the galaxies of our sample, with $N_e$ ranging from 100 to up to 2500~cm$^{-3}$. The highest values are observed at a ring at with radius of $\approx$ 1\farcs5, surrounding the nucleus. This ring seems to be correlated with the highest $\sigma$ values seen at the bottom-left/central panels of Fig.~\ref{Mrk1058}. However, the interpretation of high $N_e$ structures derived from the [S{\sc ii}] lines must be done with caution, as the relation between the line ratio and $N_e$ is rather flat for $N_e\gtrsim 2\,000$\,cm$^{-3}$ \citep{Osterbrock06} and thus estimates of $N_e$ in this range could 
also be originated by uncertainties in the line flux measurements. Within the ring, smaller $N_e$ values are seen, typically smaller than 1000~cm$^{-3}$.

\section{Discussion}\label{discussions}

\subsection{Mrk~6}

Mrk~6 is an S0a galaxy and harbors a compact Seyfert 1.5 nucleus  \citep[e.g.][]{Osterbrock76,Haniff88,Kharb06,Mingo11}.  \citet{Capetti95} presented narrow-band images of the inner 1\farcs4$\times$1\farcs4 of Mrk~6 centred at  \oiii$\lambda\lambda$4959,5007 and [O\,{\sc ii}]$\lambda\lambda$3726,29 emission-lines, obtained with the Faint Object Camera (FOC) on board the HST. Their  images show a jet-like emission feature to the south/southwest of the nucleus with extent of 0\farcs5, which is co-spatial with a radio jet seen in 6~cm MERLIN observations. 
The authors argue that these results support the interpretation that the emission is dominated by compression and heating of the gas by shocks produced by the radio jet. On larger scales, ground based \oiii$\lambda5007$ narrow-band and long slit spectroscopy at a seeing of 1\farcs0 show an extended NLR (eNLR), visible out to 35\farcs0 from the nucleus with several knots and diffuse emission, mostly to the north and south of the nucleus \citep{Kukula96}. The authors found that the eNLR is misaligned with the large scale radio axis, as revealed by the comparison of the \oiii\ image with 6 and 18 cm radio images obtained with MERLIN. 

A high spatial resolution \oiii$\lambda5007$ image of Mrk~6, obtained with the WFPC2 of HST \citep[shown in Fig.~\ref{Mrk6},][]{schmitt03} confirms the jet-like structure to the north of the nucleus (along PA = $-10^\circ$), misaligned with the major axis of the galaxy (PA = 130$^\circ$). The peak of the \oiii\ emission is observed at the nucleus and several blobs are observed around it and a  fainter emission is observed along the major axis of the galaxy \citep{schmitt03}.  

Very Large Array (VLA) observations of Mrk~6 show radio emission at different spatial scales, with bubbles observed at $\approx$ 7.5~kpc, nearly orthogonal to the inner jet that extends by approximately 3\farcs0 in the north-south direction.This complex radio emission is argued to have been possibly originated by an episodically powered precessing jet that changes its orientation \citep{Kharb06}. The inner jet -- as observed at 3.6 cm \citep{schmitt01} -- is shown over-plotted in our [O\,{\sc iii}] flux map as green contours in Fig.~\ref{Mrk6}.  X-ray observations of Mrk~6 reveal shells of X-ray emission around the radio hotspots, with a temperature of $\approx$ 0.9 keV, compatible with a scenario in which the gas in the shells is inducing a strong shock in the ISM \citep{Mingo11}.

\citet{Quillen99} present broad (F160W) and narrow (centred at the H$_2\lambda2.12\,\mu$m emission-line) band images obtained with the Near-Infrared Camera and Multi object Spectrometer (NICMOS) on board the HST. The narrow-band image does not show extended molecular hydrogen emission, attributed the nuclear source of Mrk~6  being so bright that it is difficult to observe any extended structure. 

The mass of the central SMBH is about (1 -- 2)$\times$10$^{8}$\,M$_\odot$ as estimated by Keplerian motion in the BLR using spectropolarimetry \citep{Afanasiev14} and from measurements of the 
H$\beta$ line width in combination with the reverberation lag \citep{Doroshenko12}. 

Our emission-line flux distributions (Fig.~\ref{Mrk6}) do not clearly show the jet-like compact inner emission seen in the HST image in the north-south direction due to the effect of the seeing, showing more the diffuse emission that extends to all directions. The gas kinematics shows a rotation pattern with somewhat distinct orientation in  \ha\ and \oiii.  An increase of the velocity dispersion is observed to the north in \ha\ reaching $\sigma=450$\,\kms, apparently at the tip of the nuclear jet, that could then be attributed to compression by the jet. 
The increase in $\sigma$ is lower in \oiii\ and \nii, and a possible explanation is that there is some radiation ionizing H in the plane, and this gas is compressed by the radio jet that is launched close to the plane. But the ionization axis would make an angle with the galaxy plane and higher ionization (higher \oiii/H$\beta$) would thus be observed at higher latitudes, being less affected by the compression by the radio jet. The highest $E(B-V)$ values seen at the same location could also be the result of the accumulation of dust due to compression by the radio jet. In \oiii\, there is also enhanced $\sigma$ to the west, what could be attributed to lateral expansion of the gas due to the passage of the radio jet. 
These results can be interpreted as an evidence of the interaction of gas outflows from the central AGN with surrounding gas in the galaxy, allowing the detection of emission from gas located deep in the galaxy disk (with higher extinction), and producing distortions in the gas velocity fields. 

\subsection{Mrk~79}

Mrk~79 (UGC\,3973) is an SBb galaxy harboring a Seyfert 1.2 nucleus \citep[e.g][]{Haniff88,de Vaucouleurs91,malkan98,Kraemer11}.  \citet{Peterson04} determined a mass of 52.4 $\pm$ 14.4$\times10^{6}$ M$_\odot$ for the central SMBH, through BLR emission-line reverberation.

Ground based \citep{Haniff88} and HST \citep[Top-central panel of Fig.~\ref{Mrk79},][]{schmitt03} narrow-band images in the [O\,{\sc iii}]$\lambda5007$ emission-line show a collimated emission oriented along north-south direction, extending up to $\approx$ 3\farcs0 to the north and $\approx$ 2\farcs0 to the south of the nucleus. Close to the northern edge, it bends to the to the northeast. 
Fainter emission is observed over the whole HST FoV of $8^{\prime\prime}\times8^{\prime\prime}$. \citet{schmitt03} describes the [O\,{\sc iii}] image as presenting  two blobs of emission at 0\farcs6 and 1\farcs2 south of the nucleus. To the north, blobs are seen at 0\farcs5, 0\farcs9, and 1\farcs6 from the nucleus. Considering the lower spatial resolution and lower spatial coverage of the GMOS data, as compared to the HST, our flux maps (Figs.~\ref{Mrk79} and \ref{flux_apen}) for all emission-lines are consistent with the HST image, showing a similar elongated structure along the north-south direction, with the northern side bending to the northeast and fainter emission seen over the whole GMOS FoV. 

Long-slit spectra of Mrk~79, obtained with the William Herschel Telescope (WHT), covering the wavelength ranges 3700 -- 5230\,\AA\  and 6110 -- 7460\,\AA, oriented along PA = 12$^\circ$ and PA = 50$^\circ$, reveal an eNLR at PA = 12$^\circ$, extending up to $\approx$ 15\farcs0 from the nucleus \citep{Nazarova96}. The eNLR approximately follows an asymmetric triple radio structure -- with the northern hotspot located at a distance of 800 pc from the nucleus and the southern at 460 pc from it \citep{Ulvestad84,Nagar99,schmitt01}. This is also the direction of a gas outflow, suggested by multi-components  [O\,{\sc iii}]\,$\lambda$5007 profiles along the radio axis \citep{Whittle88}. We have over-plotted contours of the radio structure as observed in 3.6 cm from \citet{schmitt01}, showing that it indeed is oriented along the extended emission observed in H$\alpha$ and [O\,{\sc iii}].

\citet{Riffel13} observed the inner 3\arcsec$\times$3\arcsec of Mrk~79 with the Gemini instrument NIFS  showing that the near-IR ionized gas emission presents a similar flux distribution to those seen in optical wavelengths. The H$_2\lambda2.12$\,$\mu$m shows, on the other hand, a more uniform flux distribution and a rotation pattern with the northwest side receding and the southeast side approaching; in addition, inflows of gas are seen along spiral arms. The ionized gas (traced by [Fe\,{\sc ii}]$\lambda1.25\,\mu$m and P$\beta$ emission) shows, besides rotation, an outflow seen as blueshifts to the north-northeast and redshifts to the opposite side. Our velocity fields are consistent with those for the ionized gas in \citet{Riffel13}: (i) we also see a rotation  component -- redshifts to the northwest, close to the border of the FoV, and blueshifts to the southeast. 
Additional support to the presence of this rotation component is given by the stellar velocity filed, that shows blueshifts and redshifts at the same locations; (ii) we also see an outflowing component -- blueshifts to the north-northeast and some (less clear) redshifts to the south-southwest, along the same orientation (PA $\approx 10^\circ$) of the outflows seen in the near-IR lines \citep{Riffel13} and at larger scales in [O\,{\sc iii}] \citep{Whittle88}. 

We note that the outflow and strongest gas emission follows the radio structure along PA = 12$^\circ$ (what we could call ``ionization axis"). One interesting result that can be seen in our data is the fact that the gas velocity dispersion is largest perpendicularly to this axis,  instead of along the direction of the outflow, a feature we have also observed in other targets \citep[e.g.][]{su96,Couto13,n5929_let,n5929,allan14b,Lena15} and that could be interpreted as due to lateral expansion of the gas due to the passage of a radio jet and/or expansion of the dusty torus surrounding the nucleus.

Regarding the gas excitation, \citet{Nazarova96} point out the observation of higher excitation along PA = 12$^\circ$, when compared to that along PA = 50$^\circ$ in long-slit spectroscopic observations along these two position angles, consistent also with our data.


\subsection{Mrk~348}

Mrk 348 (NGC\,262) is an S0/a galaxy that harbors a bright Seyfert 2 nucleus  \citep{Anton02}.  At small scales, Very Long Baseline Interferometry (VLBI) observations at 21 cm show a compact triple radio structure along PA $\approx$ 170$^\circ$, with total size of only 0\farcs2 \citep[e.g.][]{Neff83}, while at 2~cm a small-scale double source is seen with similar extent \citep{Ulvestad99}. VLA radio images at 3.6 and 20 cm show an unresolved nuclear radio source and a faint extended emission to up to 4\farcs0 from the nucleus to  the north-northeast and south-southwest, seen at 20 cm \citep{Nagar99}. \citet{falcke00} report the detection of strong H$_2$O maser emission with luminosity of 420\,L$_\odot$. 


Early ground-based, low resolution narrow-band images in [O\,{\sc iii}] and [N\,{\sc ii}]+H$\alpha$ emission suggest the presence of highly ionized region extending up to 15\farcs0 from the nucleus along the north-south direction, the same orientation of the small scale radio axis \citep{Simpson96}. 
Higher resolution HST images of the inner $7^{\prime\prime}\times7^{\prime\prime}$ also show extended [O\,{\sc iii}] emission \citep{Capetti96,Capetti02}. The WFPC2 \oiii\ narrow-band image \citep[Figure~\ref{Mrk348}][]{schmitt03} 
shows extended emission up to $\approx$ 3\farcs0 from the nucleus along PA $\approx 185^\circ$, similarly to that of  the 20~cm radio emission \citep{Nagar99}. Very close to the nucleus, the  emission is extended along PA $\approx 10^\circ$, similar to the orientation of the small scale radio jet \citep{Neff83,Ulvestad99}. Our flux maps for all emission-lines (Fig.~\ref{Mrk348} and \ref{flux_apen}) show extended emission over the whole GMOS FoV with the high intensity flux levels being more extended  along the northeast-southwest direction, in agreement with the HST \oiii\ image. Our electron density map shows two blobs of highest density  ($N_e \approx 1000$ cm$^{-3}$) that are co-spatial with the highest level emission at the nucleus and to $\approx$ 1\farcs0 to the southwest.

\citet{Fischer13} used HST STIS spectra of a sample of 53 nearby Seyfert galaxies, including Mrk~348, to study the kinematics of the NLR. They classify the \oiii\ kinematics of Mrk~348 as ``compact", as their slit position missed the extended part of the NLR. \citet{Stoklasova09} presented optical IFS of the inner $\approx$ $10^{\prime\prime}\times8^{\prime\prime}$ obtained with the OASIS spectrograph at Canada-France-Hawaii Telescope (CFHT) at an angular sampling of 0\farcs27. Their \oiii/H$\beta$ map presents higher values at the nucleus (up to 12), decreasing with the distance from the nucleus down to $\approx$ 0.5 at 3 - 4$^{\prime\prime}$ east of it, while the \nii/H$\alpha$ map shows a flat distribution of values of $\approx 1$, with a minimum seen southwest of the nucleus.  Although the range of values are similar, our line ratio maps (Fig.~\ref{Mrk348}) reveal more details due to the better spatial resolution of the GMOS data. The \oiii/H$\beta$ map shows a gradient of higher ionization to the northeast of the nucleus and lower ionization to the southwest. A similar trend is seen for the \nii/H$\alpha$ ratio, with the lowest values at $\approx$ 1\farcs0 southwest of the nucleus, at the edge of the \oiii\ blob seen in the HST image at $\approx$ 1\farcs0 to the southeast. 
Our $E(B-V)$ map shows the highest values of up to 0.7 at the nucleus. 

The gas velocity fields in \citet{Stoklasova09}, show two high velocity regions with diameters  $\approx$ 200~pc and opposite velocity signs along PA = 25$^\circ$ at distances of 300~pc from the nucleus, similarly to those observed in our velocity fields (Fig.~\ref{Mrk348}), with velocities of up to 200~\kms. These high velocity blobs are also associated to higher velocity dispersion values, of up to 200\,\kms.  \citet{Stoklasova09} interpreted these kinematics structures as corresponding to a rotating ring, inclined with respect to the galactic disc.
Our interpretation is that these two blobs originate from a nuclear outflow, due to the observation of associated higher velocity dispersion as well as to the fact that they are seen at the same orientation of the extended radio emission \citep{Nagar99}, and could thus be originated in a jet-cloud interaction. The stellar velocity field suggests a kinematic major axis approximately aligned with the photometric major axis, being almost perpendicular to the ionization axis and nulcear outflow.

\subsection{Mrk~607}

Mrk~607 (NGC\,1320) is an Sa nearly edge-on spiral galaxy hosting a Seyfert 2 nucleus \citep[e.g.][]{Mulchaey96,Ferruit00,Tsai15}, with a SMBH of mass $(5.5\pm2.5)\times10^6$\,M$_\odot$, as derived from water maser observations \citep{gao17}. Radio images at 3.6, 6 and 20 cm show a compact core with faint emission extending to the south of the nucleus, being marginally resolved in VLA observations \citep{Colbert96,Nagar99,Mundell09}. A weak radio extension is also seen to the north and northwest of the nucleus at 6 cm \citep{Colbert96}.

The \oiii$\lambda5007$ HST image \citep[top-central panel of Fig.~\ref{Mrk607},][]{schmitt03} shows extended emission by 3\farcs75 along the major axis of the galaxy  (PA = 137$^\circ$), with the highest intensity levels seen at the nucleus and to the northwest. Along the minor axis the \oiii\ emission extends by 1\farcs35. A similar flux distribution is seen in an \nii+H$\alpha$ HST image \citep{Ferruit00}. Large scale narrow-band images for the  H$\alpha$+\nii\ and [O\,{\sc iii}]$\lambda5007$ lines show extended emission up to $\approx$ 15\farcs0 from the nucleus along the major axis of the galaxy \citep{Mulchaey96}. Our flux maps (Figs.~\ref{Mrk607} and \ref{flux_apen}) show emission over the whole GMOS-IFU FoV, with the highest intensity levels along the major axis of the galaxy with some enhancement towards the northeast (where the extended HST [O\,{\sc iii}] emission is mostly observed) in agreement with previous \oiii\ and H$\alpha$+\nii\ images \citep{Mulchaey96,schmitt03,Ferruit00}. 

\citet{Ferruit00} present an \oiii$\lambda5007$/(H$\alpha$+\nii) ratio map, which shows the highest values at the nucleus and smaller values to the northwest. This map can be compared with our \oiii$\lambda5007$/H$\beta$ (Fig.~\ref{Mrk607}) map, which shows high values of up to 10 at the nucleus and to the northeast, decreasing a bit to the south, and  smaller values to the northwest, as observed by \citet{Ferruit00}.  

Our $E(B-V)$ map  for Mrk~607 shows low values to the northwest, indicating that the collimated line emission seen at this orientation arises from a low-extinction gas. In addition, a trend of higher $E(B-V)$ values to the northeast and lower values to the southwest, along the minor axis of the galaxy is in agreement with large scale $B-I$ colour maps of Mrk~607, that show higher extinction to the northeast of the nucleus, indicating that the northeast is the near side of the galaxy disk \citep{Kotilainen98}.

The gas velocity fields (Fig.~\ref{Mrk607})  suggest ordered rotation  with the northwest side approaching and the southeast side receding. The stars, on the other hand, rotate in the opposite direction, with blueshifts to the southeast and redshifts to the northwest, in good agreement with the stellar velocity field presented in \citet{Riffel17}, derived from the fitting of the K-band CO absorption band heads using Gemini NIFS spectra. The opposite directions of the gas and stars rotation may be related to gravitational interaction of Mrk\,607 and its companion NGC\,1321 \citep[e.g.][]{hunt99}.  
The $\sigma$ maps for the gas show the highest values of up to 130\,\kms\ approximately along the minor axis of the galaxy, while the stellar $\sigma$ map presents predominantly small values  ($\approx 50$\,\kms) at these locations. As the ionization axis seems to be perpendicular to these structures of high gas $\sigma$, one possibility is that they are tracing equatorial outflows as observed for other active galaxies \citep[e.g.][]{Couto13,allan14b,n5929_let,Lena15}, giving support to models of equatorial accretion disk winds \citep{li13} and outflowing torus  \citep{honig13,elitzur12,ivezic10,mor09,nenkova08,elitzur06}. The origin of these high gas $\sigma$ structures cannot be attributed to gas located in the disk, as the stellar velocity dispersion maps present small values at these locations. However, a detailed analysis of the gas kinematics is needed to get a final answer on the origin of these high $\sigma$ structures.

\subsection{Mrk~1058}

Mrk~1058 is an isolated spiral galaxy, classified as Sb and although it harbors a bright Seyfert 2 nucleus \citep[e.g.][]{Corwin94,malkan98,Chapelon99,Smirnova10}, it lacks detailed studies in the literature.  Broad band HST images reveal dust lanes to the north of the nucleus, suggesting this is the near side of the galaxy \citep{malkan98}.

According to \citet{DeRobertis86}, an integrated nuclear spectrum (aperture 2\farcs7$\times$4\farcs0) shows weak emission in H\,{\sc i}  and  He\,{\sc ii} recombination lines and [O\,{\sc i}]. They reported also that the [N\,{\sc ii}] lines are strong relative to H$\alpha$, a result confirmed by our BPT diagram (Fig.~\ref{bpt}) that shows that Mrk~1058 presents the highest average values of  [N\,{\sc ii}]$\lambda6583$/H$\alpha$ of our sample. In addition the   [O\,{\sc iii}]$\lambda5007$/H$\beta$ intensity ratio presents high values at all locations.

Mrk~1058 presents only a faint nuclear radio emission at 3.6 cm, with no detected extended emission in  VLA observations \citep{Kinney00,schmitt01}. The HST \oiii$\lambda$5007 image \citep[Figure~\ref{Mrk1058},][]{schmitt03} shows extended emission to up to 2\farcs0 to the southwest, with a  V-shaped morphology with opening angle of 55$^\circ$ with central axis oriented along PA = 205$^{\circ}$, perpendicular to the major axis of the galaxy \citep{schmitt03}. Our \oiii\ flux map shows emission over the whole GMOS-IFU FoV, but enhanced emission is indeed observed towards the southwest, although not resolving the structure seen in the HST image due to the lower resolution.


The \ha\ flux distribution is somewhat distinct from that of \oiii\, showing extended emission to the northwest and southeast of the nucleus. At the edge of these structures (at $\approx$ 2\farcs0 northeast and $\approx$ 2\farcs0 southwest) smaller values of \nii/H$_\alpha$ are observed. One possible interpretation is the presence of regions of star formation at these locations. Due to the low-intensity (or non detection) of the H$\beta$ line, we were able to construct the [O\,{\sc iii}]/H$\beta$ map for these regions only for the inner 1\farcs5 radius. In this region, there is an ionization gradient, with highest values to the west and lowest to the east.


As for the flux distributions, the kinematics for H$\alpha$ and \oiii\ are also distinct from each other, with the H$\alpha$ velocity field showing a distorted rotation pattern while in the case of \oiii\ this pattern is less clear. Both the \ha\ and  \oiii\  velocity fields seem to show additional blueshifts to the southwest, coincident with the elongated emission seen in the HST \oiii\  image. The velocity dispersion maps show also enhanced values coincident with the blueshifted region. The stellar velocity field also shows a somewhat similar rotation pattern to that of the gas but does not show the blueshift to the southwest. Considering all these
 these characteristics, combined with the ionization gradient discussed above, we interpret these blueshifts as being originated in gas outflowing from the nucleus.  

One particular characteristic of this galaxy is the enhancement of gas density towards the borders of the field at about 1\farcs7 from the nucleus, that seems to coincide with regions of enhanced velocity dispersion in [O\,{\sc iii}]. We speculate that these enhancements are produced by the outflow pushing the surrounding medium, but the uncertainties in $N_e$ may be high at these locations as the relation between the [S\,{\sc ii}] line ratio and $N_e$ is rather flat for such high values \citep{Osterbrock06} and thus high  $N_e$ values could also be produced by uncertainties in the [S\,{\sc ii}] line flux measurements.



	



\section{Conclusions}

We have used GMOS-IFU observations covering the wavelength range from 4300\,\AA\ to 7100\,\AA\  to map the emission-line flux distributions and kinematics,  as well as the stellar kinematics,  of the inner kiloparsec of five nearby Seyfert galaxies -- Mrk\,6, 79, 348, 607 and 1058 --  at a spatial resolutions ranging from 110 to 280 pc. In this first paper we present gas flux, excitation, and kinematic maps, reddening and density maps, as well as stellar kinematic maps. In a forthcoming paper we will use these measurements to obtain gas masses, to model the gas velocity fields in more detail, and to estimate mass flow rates.

The main results of this paper are:

\begin{itemize}

\item Extended emission over the whole GMOS-IFU FoV (3\farcs5$\times$5\farcs0, corresponding to the inner 1 -- 3 kpc at the galaxies) is observed for the strongest emission-lines: H$\beta$, \oiii$\lambda$5007, H$\alpha$,  \nii$\lambda6583$ and \sii$\lambda\lambda$6716,31. The \oiii$\lambda5007$/H$\beta$ vs. \nii$\lambda6583$/H$\alpha$ diagnostic diagram is consistent with the line emission being originated from gas excited by the central AGN for all galaxies, although gradients of these line ratios are observed in flux-ratio maps.

\item The highest reddening ($E(B-V)$ $\approx$ 1) is usually observed within the inner few hundred pc around the nucleus and/or at regions with the highest excitation.

\item The highest gas densities ($N_{e} \approx$ 1000 -- 2000 cm$^{-3}$) are usually observed at the nucleus, in a few cases also extending towards regions of highest excitation. A particular case is Mrk\,1058 that seems to show a circumnuclear ring of high density gas at $\approx$ 1\farcs7 (592 pc) from the nucleus.

\item The gas kinematics show a distorted rotation pattern that can be attributed to a combination of emission from gas in  rotation in the galaxy plane and outflows.

\item The rotation component of the gas is confirmed by their observation in the stellar kinematics,  except for Mrk\,607 in which the gas is counter-rotating relative to the stars.

\item The gas velocity dispersion shows two typical patterns: it is enhanced at the location of the outflows or at the nucleus perpendicularly to the outflow. This latter behavior has been attributed to an equatorial outflow at the AGN, possibly originating in the torus.

\item The (projected)  velocities of the outflow reach at most $\approx$ 200\,km\,s$^{-1}$, but could be larger as they seem to be mostly in the plane of the sky. The apparent geometries of the outflows can be described as follows:

(i) Mrk\,6 shows enhanced gas $\sigma$ values surrounding the radio jet and distortions in the velocity field interpreted as due to an outflow along the jet, but the geometry needs further constraints from modeling.

(ii) Mrk\,79 shows a bipolar outflow oriented along the ionization axis inferred from the HST \oiii\ image; another interesting feature is enhanced velocity dispersion in a nuclear strip perpendicular to the outflow that we interpret as due to lateral expansion produced by the passage of the radio jet.

(iii) Mrk\,348 shows a very clear bipolar outflow in two blobs with opposite velocities relative to the nucleus and enhanced velocity dispersion.

(iv) Mrk\,607 is seen close to edge on, with the ionization axis along the galaxy plane, and shows distortions in the velocity field that could be due to an outflow that needs further constraints from modeling; enhanced velocity dispersion perpendicular to the inferred axis of the ionization cone and radio axis could be interpreted as lateral expansion of the gas as in the case of Mrk\,79.

(v) Mrk\,1058, besides presenting a rotation pattern seen in H$\alpha$ shows also blueshifts mostly in \oiii\ in association to enhanced velocity dispersion and extended \oiii\ emission that can be interpreted as the blueshifted part of a nuclear outflow.

\end{itemize}

In summary, we have found signature of outflows in all galaxies, but their geometry, intrinsic velocities, and resulting mass flow rates will be presented in a forthcoming study based on multi-components fits to the emission-lines.

\section*{Acknowledgments}
We thank an anonymous referee for his/her thorough review, comments and
suggestions, which helped us to significantly improve this paper.
Based on observations obtained at the Gemini Observatory, which is operated by the Association of Universities for Research in Astronomy, Inc., under a cooperative agreement with the NSF on behalf of the Gemini partnership: the National Science Foundation (United States), the National Research Council (Canada), CONICYT (Chile), Ministerio de Ciencia, Tecnolog\'{i}a e Innovaci\'{o}n Productiva (Argentina), and Minist\'{e}rio da Ci\^{e}ncia, Tecnologia e Inova\c{c}\~{a}o (Brazil). I.C.F.  thanks the financial support received from CAPES.   
R.A.R. acknowledges support from FAPERGS and CNPq.

\appendix

\section{Emission-line flux distributions}\label{apen}
In this section we present the flux distributions 
 H$\beta$, [O\,{\sc i}]\,$\lambda$6300, [N\,{\sc ii}]\,$\lambda$6583 and [S\,{\sc ii}]\,$\lambda$6731 emission-lines for Mrk\,6, Mrk\,79, Mrk\,348, Mrk\,607 and Mrk\,1058.

\begin{figure*}
\centering
   \includegraphics[scale=0.65]{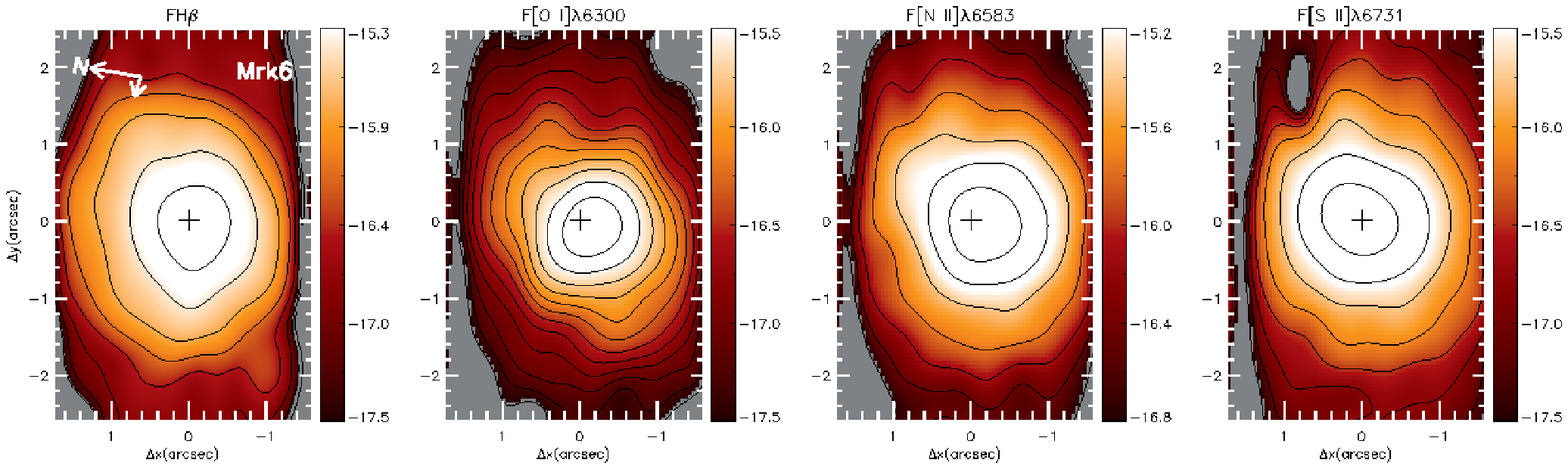}
      \includegraphics[scale=0.65]{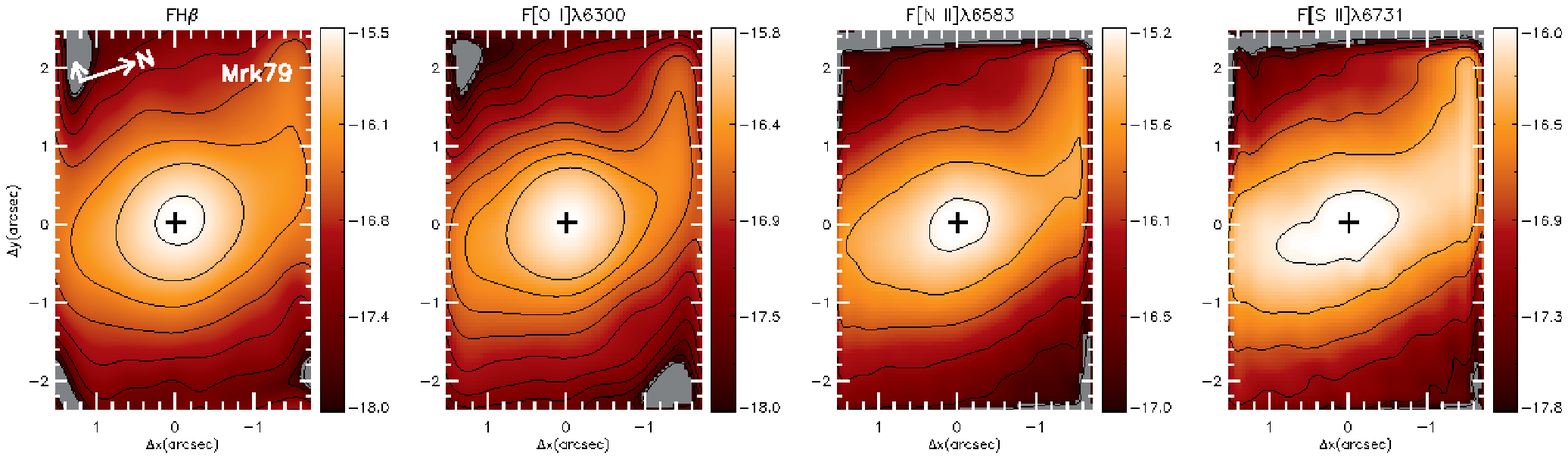}
   \includegraphics[scale=0.65]{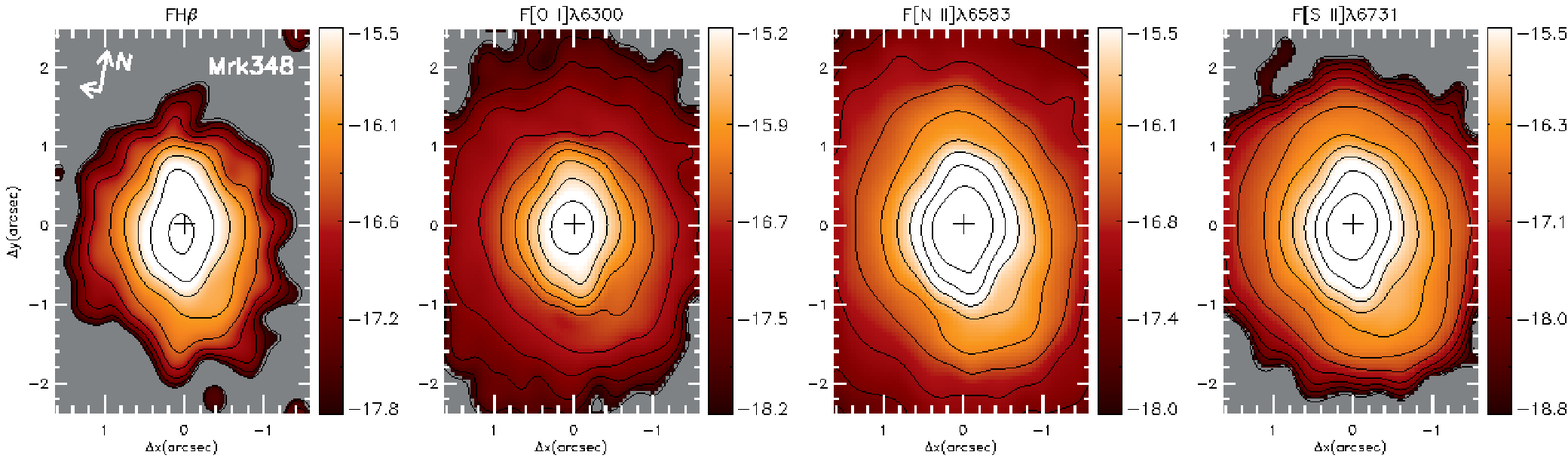}
   \includegraphics[scale=0.65]{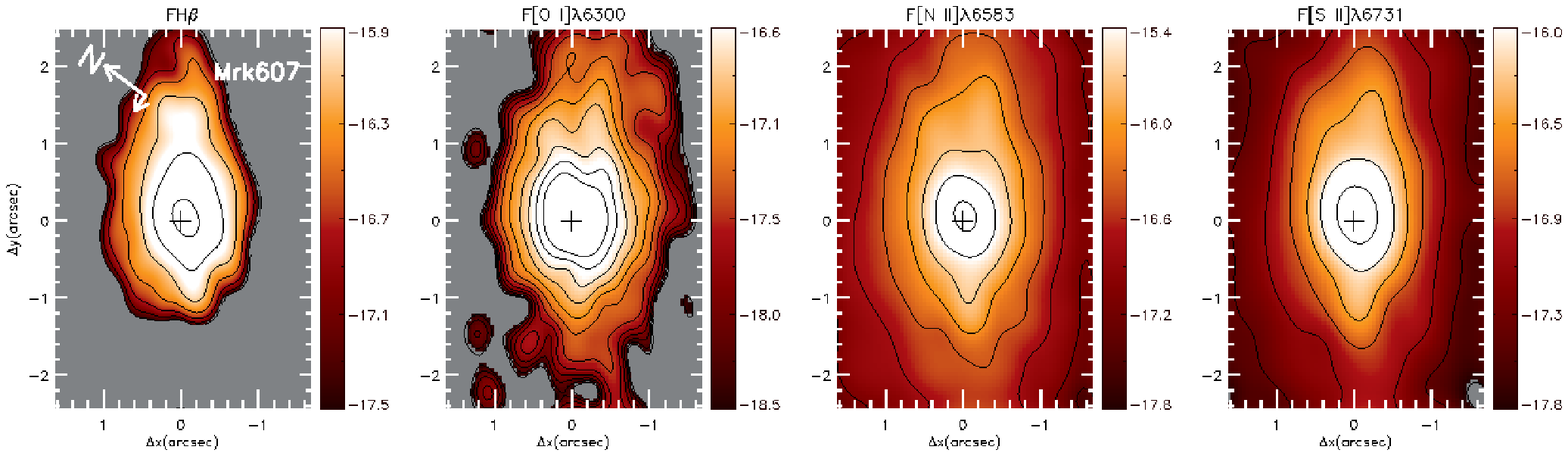}
      \includegraphics[scale=0.65]{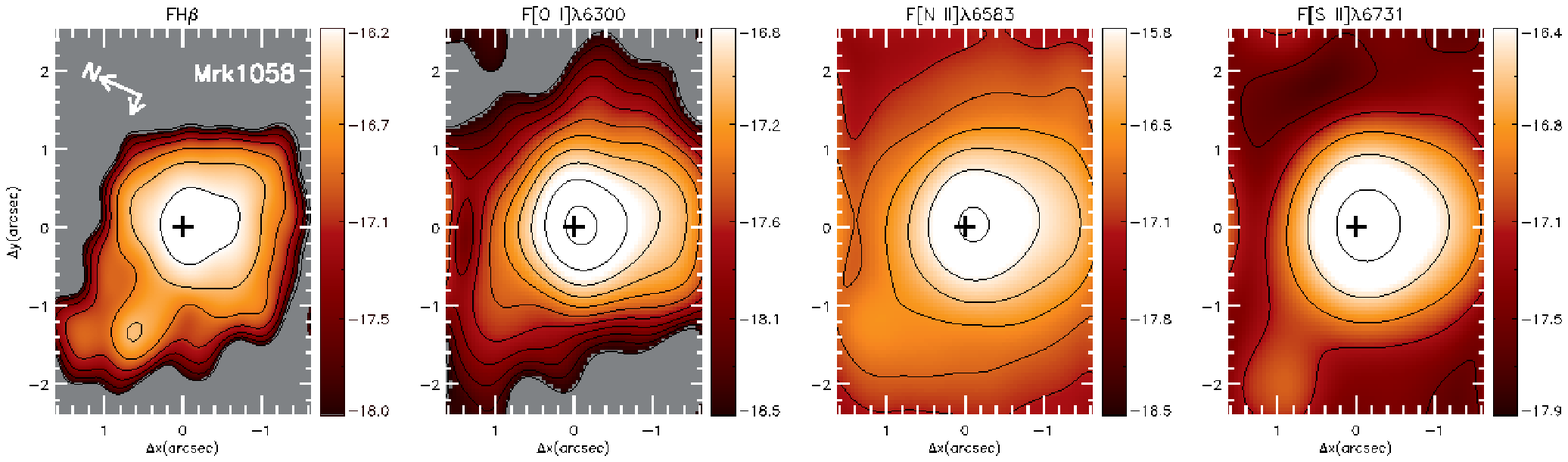}
     \caption{Flux distributions for H$\beta$, [O\,{\sc i}]\,$\lambda$6300\,\AA\, [N\,{\sc ii}]\,$\lambda$6583\,\AA\ and [S\,{\sc ii}]\,$\lambda$6731\,\AA\ emission-lines. The color bars show the flux scale in logarithmic units of erg s$^{-1}$cm$^{-2}$. From top to bottom: Mrk\,6, Mrk\,79, Mrk\,348, Mrk\,607 and Mrk\,1058.}
\label{flux_apen}
\end{figure*}

\label{lastpage}

\end{document}